\documentclass[fleqn,referee,epsfig,color]{aastex63}

\usepackage[T1]{fontenc}
\usepackage{ae,aecompl}

\usepackage{graphicx}	
\usepackage{amsmath}	
\usepackage{amssymb}	

\newif\ifAMStwofonts
\newcommand{\be}{\begin{equation}}
\newcommand{\ee}{\end{equation}}
\newcommand{\bea}{\begin{eqnarray}}
\newcommand{\eea}{\end{eqnarray}}

\newcommand{\W}{{\bf W}}

\received{}
\revised{}
\accepted{}
\submitjournal{ApJ}

\shorttitle{Galactic Magnetic Dynamos}
\shortauthors{Henriksen \& Irwin}

\begin{document}

\title{\bf  Turbulent Magnetic Dynamos with Halo Lags, Winds, and  Jets }

\correspondingauthor{R. N. Henriksen}
\email{henrikr@queensu.ca}

\author{R. N. Henriksen \& Judith Irwin}
\affiliation{Dept. of Physics, Engineering Physics, \& Astronomy \\
Queen's University \\
Kingston, Ontario, K7L 3N6, Canada}



\begin{abstract}
This paper presents scale invariant/self-similar galactic magnetic dynamo models based on the classic equations, and compares them qualitatively to recently observed magnetic fields in edge-on spiral galaxies.
We classify the  axially symmetric dynamo magnetic field  by its  separate sources, advected flux and sub scale turbulence. We neglect the diffusion term under plausible physical conditions.
 There is a time dependence determined by globally conserved  quantities.
 We show that magnetic scale heights increase with radius and wind velocity.   We suggest that AGN outflow is an important element of the large scale galactic dynamo, based on the dynamo action of increasing sub scale vorticity. This leads us to {\it predict} a correlation between  the morphology of coherent  galactic magnetic field (i.e. extended polarized flux) and the presence of an AGN.   

\end{abstract}

\section{Introduction and Background}
Dynamo theory as an explanation  for the magnetic fields in galaxies, including the Milky Way, has a long history both observationally and theoretically. Observational reviews  and discussion of the relevant theory may be found in  \cite{Beck2015} and \cite{Kr2015}. The textbook \cite{KF2015} discusses the theory  and observation with abundant references.  \cite{Black2015} discusses modern theoretical developments  in the theory, although these are not employed in our work. 

More recently, a scale invariant version of the classical theory
\citep[e.g.][]{SKR1966,M1978,Hen2017,HWI2018a,HWI2018b,WHIM-P2019} has had some success. 
This approach succeeds in isolating and predicting the more common qualitative elements of a galactic magnetic field, much of which is compatible with previous work. Observational data have been enhanced by the CHANG-ES survey, which has been reviewed recently in \cite{CHANGESXVIII}.  The presence of magnetic  spiral `arms' or `ropes'  in the radio halos of galaxies was  predicted originally by this approach in \cite{Hen2017}  and subsequent   work by this author and colleagues. A key prediction has been supported recently by the inferred magnetic field in the radio halo of the  `Whale' galaxy NGC~4631, as described in \cite{CHANGESXV} and \cite{WHIM-P2019}.\footnote{For a visualization see also {\it https://public.nrao.edu/news/giant-magnetic-ropes}}

 In parallel to the observational/analytic studies,  numerical work has blossomed. This approach studies either the cosmological origin  of galactic magnetic fields, or the  development from a `seed field'  of a galactic dynamo. We do not pretend  to review this literature fairly, but we can refer to some that have been brought to our attention. 

In the cosmological category we find early work by \cite{WA2009} and \cite{BeckAM2012}, and this has been continued into the present  by \cite{PMS2014}, \cite{RT2016}, \cite{RT2017a}, \cite{RT2017b}, and \cite{PGP2018}, among others. {The  paper, \cite{RT2016}, is noteworthy for establishing that global, supernovae driven, turbulence during galaxy formation  can amplify seed fields to necessary low red shift values. This is really a globally distributed  $\alpha/\Omega$ dynamo, in which one finds limiting exponential field growth with time scale $1/\Omega$.  Our limiting time dependence found below under the assumption of scale invariance agrees with this behaviour.} In \cite{KMD2019} and  \cite{MSD2020} the authors  introduce the notion of tracking the cosmological history of a given low Z magnetic field. 

The cosmological environment, beyond  clusters, is not suitable for the assumption of  asymptotic scale invariance because of large scale continuing evolution. The extended infall onto a forming galaxy is a possible exception.   However, the numerical work that focuses on the origin and structure of galactic fields should, when smoothed appropriately, correspond to our assumption of self-similarity (including similarity between galaxies). Recent papers  in this category by \cite{BZK2017} and \cite{PVB2020} are particularly relevant. We will compare our results with theirs in a later discussion after we have presented our calculations and observations. 

{There is a growing consensus that galactic winds are an essential component of the long lived galactic dynamo. In fact throughout this series of papers on scale invariance, we have found it necessary to include such winds. This was already the case in our original suggestion \cite{HI2016}, which required halo lag and disc wind to develop an X magnetic field. In the current paper we also call attention to AGN outflows as being possibly a part of the dynamo \citep[see also][]{PSB2020}.
 The numerical work attempts to reveal the detailed origins of such winds, whereas this paper does not make statements about the wind origin.  Rather, we take the wind to be a necessary part of the self-similar, scale-invariant development which we present below. }

{Recently, several relevant numerical/analytic works  have been submitted in draft form to  ariv.org. Two papers, \cite{QJT2021a}, \cite{QTJ2021b},
have examined galactic  winds driven by cosmic rays, either by random diffusion or by streaming along the magnetic field. It appears that diffusion is more effective, but it is not clear that such winds  can produce the observed X field at all heights above the disc. An alternate approach in a recent draft, \cite{SDL2021}, suggests magnetic pressure can drive a disc outflow, after the field has been amplified by the $\alpha/\Omega$ dynamo. The amplification in the disc is much like that proposed in \cite{BZK2017}.  This paper includes the outflow as an integral part of the dynamo. Moreover the paper presents a biconical outflow from the galaxy that might coincide with the  {\it large scale} soft X-ray emission found in \cite{PSB2020}. }
  
  {The more traditional wind  from the disc is driven by vigorous star formation with the consequent super nova explosions. This has not seemed to be efficient enough, but there is a recent, very relevant paper \cite{SHH2018}. This paper shows in numerical detail how discrete localized supernovae enhance the outflow over what is achieved by smoothly averaging the super nova energy input over the star forming disc.}

{Recent numerical papers \cite{GPS2021}, and \cite{ATK2021} have addressed the problem of primordial magneto genesis. These study the development of primordial magnetic fields at ionization fronts during the recombination due to the well known Biermann battery and a recent mechanism called the Durrive battery.  They follow the field and galaxy evolution  to the current epoch and find \citep{GPS2021} that the different initial field seedings have small ultimate effect. This may be due to the turbulent dynamo occurring naturally and becoming dominant, as found in \cite{RT2016}.}

The application of scale invariance/self-similarity to such complicated physical problems, relies on the whole system of interest attaining an internal asymptotic behaviour. The philosophy is similar to applying thermodynamic laws to the beginning and end of a string of chemical reactions  without the intervening details.  The numerical work furnishes the physical details at a cost of complexity, while the scale invariance/self-similarity yields only ultimate behaviour, although  it is frequently exact. 

 Our theoretical technique is the same as always in this series, but we have neglected intermediate scale diffusion in favour of other parts of the dynamo, namely small scale (`fast') turbulence and outflow/inflow. This allows simpler analytic results for the most part. Halo lag and `fast turbulence' throughout the disc and halo are new parts of the assumed self-similarity, as is the possible contribution of  an AGN to the dynamo. 
 
 A slightly different approach to understanding the magnetic field in edge-on galaxies has been pioneered, for example, in \cite{FT2014} and \cite{TF2017}. In this approach, different topological classes of magnetic field are assigned to a galaxy, and then the consequent pattern of Faraday Rotation is carefully calculated. This allows comparison with observations, particularly for the Milky way. These fields are somewhat arbitrary, but in fact they correspond quite closely to our scale invariant dynamo fields.      

{\bf \subsection{Qualitative Model and Philosophy}}

It helps when interpreting the  magnetic field topology that we find,  to recall that 
{\it scale invariant fields must appear self-similar at all resolutions and so they normally only close at infinity.} This is true even though the divergence of the field is everywhere zero, so that a field line is never broken. Exceptional cases do arise wherein the field lines close through a  singularity,  but we do not study these here.

In this paper we study the structure of scale invariant  (or self-similar)  magnetic dynamo fields, which contain,  for the first time, both a rotating halo that lags behind the disc rotation and {an} accelerating outflow/inflow.  Observational examples of halos that are best fit with accelerating outflows are NGC~3556 \citep{misk19} NGC~891 \citep{schmi19},  and NGC~5775 \citep{heal21}. We also allow the (scaled) sub scale  (also called `fast' elsewhere) turbulent vorticity (sometimes described as hydrodynamic helicity)  to increase with latitude in an attempt to imitate an AGN. The usual scale invariant variation of this quantity is proportional to cylindrical radius, and hence it decreases near the axis.

The `small' (i.e. sub resolution scale) scale turbulence must have net hydrodynamic vorticity for the $\alpha$ effect to be present, so we often write sub scale vorticity as synonymous with the  $\alpha$ effect. Because of a notational clash with our temporal scale factor $\alpha$, we refer to the effect as the $\alpha_h$ effect.

 
 The inclusion of a  rotating galactic halo that lags the disc is an important observed feature that we include in the dynamo action.
 Rotational velocity in the galactic halo gas  significantly declines with height above the disc of spiral  galaxies 
\citep[e.g.][]{R2000,H2007}.
 This `halo lag'  inspired  a model \citep[][= H-I]{HI2016} wherein the halo magnetic field is generated through  coupling to the distant halo and/or the intergalactic medium.  This model  accounts for both the lagging  halo rotation and a global magnetic field. The model is dynamical in the sense that pressure and gravity  also contribute to the steady state, but it did not include the sub scale turbulent dynamo.


The present formalism follows that in \cite{HWI2018a} so the formalism is only briefly introduced in the next section. The scaled lag and wind/accretion are allowed  to be variable with the latitude in the halo. However {we do} not  solve for the dynamics of the halo gas. Rather, according to our basic assumption, the halo lag, outflow/inflow, and sub scale helicity are determined in functional form by the scaling symmetry. The time dependence is  determined by the global invariants under the scaling symmetry, much as  the scale invariant dissipation rate determines the  Kolmogorov scaling law in hydrodynamic turbulence.  

The  H-I model led to a rather coherent field structure. The celebrated `X-type' projected magnetic field 
\citep{Kr2015,Beck2015}
was present, as were also spiralling field lines (axially symmetric) rising  and expanding into the galactic halo. 
In this paper we wish to determine {\it whether there are characteristic observable  galactic magnetic topologies that may distinguish between a dominant $\alpha_h$ effect or a dominant macroscopic dynamo}.  We compare some of our results to the recent catalogue of magnetic fields in the CHANG-ES survey given in \cite{KIS2020}. All of these images are freely available for download in FITS format at queensu.ca/changes.

 \vskip 0.2truein

\section{ Basic Theory}
\label{sect:basic}

We refer to \cite{HWI2018a}  and references therein for a  detailed presentation of  scale invariant or self similar classical dynamo theory, including time dependence and axial symmetry.  However in this study we set  the diffusivity equal to zero, but include a lag in the rotational halo velocity as well as accelerating/decelerating  outflow or accretion, from or  onto the disc. An innovation is that the sub-scale  turbulent helicity is allowed to vary with height in addition to the usual variation with radius. This produces an axial magnetic field that may in reality be generated by jets or nuclear winds.

It is worth discussing briefly the conditions under which we may justifiably  neglect the diffusion term in the dynamo equation (Eq.~\ref{eq:dynamo1}) below. If we compare in order of magnitude  the terms in  the diffusion coefficient $\eta$ and  in the sub scale helicity $\alpha_h$  to the convection term we obtain respectively the ratios 
\be
\frac{\eta}{LV},~~~~~\frac{\alpha_h}{V},
\ee
where $L$ is a macroscopic spatial scale and $V$ is a macroscopic velocity. One estimate for the diffusion coefficient, $\eta$, is the turbulent value, $u\ell$, where $\ell$ is the sub scale and $u$ is the corresponding sub scale turbulent velocity. We will take this to be  $u\approx \alpha_h$. Then the first ratio is small in proportion to $(\alpha_h/V)(\ell/L)$.  It is small compared to the second $\alpha_h$ term insofar as $\ell/L$ is small. In the CHANG-ES data, a linear resolution scale on a galaxy is typically several hundred parsecs, which may be regarded as  a minimum $L$. The sub scale  is on the scale of star formation so that $\ell/L$ is $O(10^{-2})-O(10^{-1})$. Similarly $u/V$ will be small, if $u$ is a turbulent velocity and $V$ is macroscopic.

Another estimate of $\eta$ is  as a Bohm type value due to gyrating  relativistic electrons in a magnetic field. This suggests
\be
\eta\approx \frac{cE_e}{3eB},\label{eq:Bohm}
\ee
where $E_e$ is the electron total energy and $e$ is the electronic charge in esu. This gives $\eta\approx 1.7\times 10^{13}\gamma_e/B$ cgs, where $\gamma_e$ is the electron Lorentz factor. For $B\approx 10^{-6}$ Gauss and $\gamma_e\approx 10^{5}$ we obtain $1.7\times 10^{24}$ cgs. This is to be compared to $LV$ which is $10^{27}$ cgs for $L\approx 10^{20}$ cm and $V\approx 10^7$ cm/s. We conclude that  the diffusion is a higher order effect in the general dynamo in the presence of convection.

 Moreover, in previous work \cite{Hen2017}, \cite{HWI2018a}, we have included the intermediate scale diffusion in our self-similar solutions. These solutions are generally more complicated to obtain, but they do not show qualitative differences with the solutions in this paper. We believe that this is because the mean outflow or inflow dominates the intermediate scale diffusion. 


It is difficult to reconcile the kinematic scale invariant symmetry that we assume with  the  actual observed halo lag (although see H-I).  Strictly, a full treatment would require multi dimensional self similarity  and  so plunge the problem back into partial differential equations (although without the time dimension \cite{CH1991}). We achieve a reasonable form in this kinematical study with one self-similar variable, by assuming  that the dynamo equation  holds in a pattern reference frame rotating with the velocity ${\bf v}_p=v_p(r) \hat {\bf e}_\phi$. Here $e_\phi$ is the azimuthal unit vector around the rotation axis of the galaxy. 

A  valid classical dynamo equation in the pattern frame  requires Faraday's law  to hold  plus the  usual dynamo expression for the  electric field \citep[e.g.][]{Hen2017}. One finds, starting with Faraday's law in the systemic frame (assumed inertial), that to first order in $v/c$ \footnote{Hence to this order  the magnetic field is invariant between the systemic frame and the rotating pattern frame.}  Faraday's equation does hold  in the pattern frame {\it provided that the pattern speed is solid body rotation}.  A solid body rotation does not itself contribute to dynamo action. This permits the magnetic field to be invariant on transforming to this frame of reference. Unfortunately such a uniform rotation does not correspond to the principal rotation of the galaxy. Rather it corresponds to a spiral arm pattern rotation or perhaps the pattern rotation of magnetic spiral arms. {\it Nevertheless the invariance between the pattern frame and the inertial systemic frame means that the observed field should correspond to that found in the pattern frame}.

Our total rotational speed will be $v_p+v_s(z)$, where $v_s(z)$ is the  halo rotational velocity  in the pattern frame, normally a lag relative to $v_p$. It would be ideal to have $v_p$ constant, if we identify the pattern speed with the galactic thin disc rotation. Unfortunately this choice adds a spurious first order term to the pattern frame Faraday's equation, namely $-v_pB_r/(rc)$ added to the time derivative. This term represents the dynamo action of a constant rotation velocity.  It may only be neglected compared to the time derivative if our vertical lag rate (that is  $\delta$ below) is larger that $v_p/r$.  


We proceed to solve the classical dynamo  equation  
\citep{SKR1966,M1978}
in the pattern frame for the magnetic vector potential  (we set $\eta=0$  which is included in the following equation for reference only)

\be
\partial_t{\bf A}=\alpha_h\nabla\times{\bf A}+{\bf v}\times \nabla\times {\bf A} -\eta\nabla\times\nabla \times {\bf A},\label{eq:dynamo1}
\ee
where a scaled magnetic field  and the scaled vector potential are related by 
\be
{\bf b}=\nabla\times {\bf A},\label{eq:B}
\ee
 $\alpha_h$ is the sub-scale turbulent `helicity', and ${\bf v}$ is the global vector velocity of the halo gas.

 For Dimensional\footnote{Capital D is used to distinguish a degree of freedom in parameter space from a geometrical dimension.} simplicity, the scaled magnetic field is taken as  
 \be
 {\bf b}\equiv \frac{{\bf B}}{\sqrt{4\pi\rho}},\label{eq:b}
 \ee
 where $\rho$ is some arbitrary constant with the Dimension of density. This gives ${\bf b}$ the Dimension of velocity and so ${\bf A}$ has the Dimension of specific angular momentum.
 The formula for the vector potential requires neglecting an electrostatic field that would result from large scale charge separation after integration of the Faraday's law.   It seems unlikely that a large scale electrostatic field should be important. Such an integration has the additional benefit of allowing the sub scale helicity $\alpha_h$ (and the diffusion coefficient $\eta$ where applicable) to vary with position without adding extra terms.




The search for  axially symmetric (we use components in cylindrical coordinates)\footnote{There is no consequent axial symmetry in the  Cartesian field components. Absolute axial symmetry requires a coordinate invariant definition.}, scale invariant solutions of Eqn.~\ref{eq:dynamo1} requires assuming that 
\be
{\bf A}=\bar{\bf A}(R,Z) e^{(2\delta-\alpha)T},\label{eq:A1}
\ee
where the scale invariants  (equivalent to the self-similar variables ) in space and time are 
\be
R\equiv re^{-\delta T},~~~~~Z\equiv z e^{-\delta T},~~~~~e^{\alpha T}\equiv (1+\alpha t).\label{eq:invariants}
\ee
As in \cite{HWI2018a} the reciprocal scales in space and times are $\delta$ and $\alpha$, respectively. If there were a fixed spatial scale in the problem then $\delta=0$, while a fixed time scale requires $\alpha=0$. In the latter case our time, $T$, would be defined with $\delta$ replacing $\alpha$. The limit $\alpha=0$ is not strictly equivalent to an exact steady state 
\citep[e.g.][and below]{Hen2017,HWI2018a}. Extensive discussion of the use of such scales may be found in \cite{Hen2015}, but in a given problem they may be assigned to give numerical quantities. 

The other quantities  must have the following forms relative to the pattern frame
\bea
{\bf v}&=&\tilde\alpha_d\delta R e^{(\delta-\alpha)T}(0,v,w),\label{eq:velocity}\\
\alpha_h&=& \tilde\alpha_h\delta R e^{(\delta-\alpha)T},\label{eq:helicity}\\
{\bf b}&=& \bar b(R,Z) e^{(\delta-\alpha)T},\label{eq:field}
\eea
where we have allowed  only an azimuthal velocity  and an outflow/inflow  wind component in this study. At this stage the auxiliary functions  $\tilde\alpha_h$ and $v$, $w$ may be arbitrary functions of $R$ and $Z$. 
 
The similarity class 
\citep[$a$, see e.g.][]{CH1991,Hen2015}
is sometimes used in what follows, and  it is given in terms of the ratio of characteristic temporal and spatial scales according to 
\be
a\equiv \frac{\alpha}{\delta}.\label{eq:a}
\ee

 To this point  we have described a {\it multi-variable} formulation of scale invariance (the invariant scaling follows by holding the self-similar variables constant), wherein the time dependence has been reduced to either a multiplication by a power law or by an exponential, depending on the class $a$. Proceeding with  this analysis  requires the solution of a partial differential equation  in $R$ and $Z$, once the auxiliary functions are given in the same variables. However, inspired mainly by the X-type field structure \citep[e.g.][]{KIS2020}, we reduce the problem further in this series of papers by looking for a solution that depends only on the tangent of the  halo latitude, in the form
 \be
 \zeta\equiv\frac{Z}{R}\equiv \frac{z}{r}.\label{eq:zeta}
 \ee
Such a solution is found to exist provided that $\tilde\alpha_h$ and $v$ and $w$ are at most functions of $\zeta$ and now 
\be
\bar{\bf A}(R,Z)=\tilde {\bf A}(\zeta),~~~~~\bar{\bf b}(R,Z)=\frac{\tilde{\bf b}(\zeta)}{R}.\label{eq:A2b2}
\ee

For our halo lag description we take the Dimensionless velocity in the pattern frame (referred to as $v_s$ earlier)
\be
v(\zeta)=-v_1|\zeta| \label{eq:v}
\ee
for constant $v_1$.
With this choice  the explicit rotational velocity  lag  and wind/accretion velocity in this frame, by Eqs.~\ref{eq:velocity} and \ref{eq:invariants}, become respectively 
\be
v_\phi=-\tilde\alpha_h\delta(v_1|z|) e^{-\alpha T},~~~~v_z=\tilde\alpha_h w(\zeta)\delta r e^{-\delta T}.\label{eq:vphi}
\ee
When $\tilde\alpha_h$ is constant, the halo lag, $v_\phi$, is linear in $z$, but should $\tilde\alpha_h$ be  linear in $\zeta$  then the the lag is proportional to $z^2/r$ (see also  Sect.~\ref{sect:aux3}).
The consequent systemic form after adding a constant pattern velocity, $v_p$ , is similar to that found dynamically in \cite{HI2016}. The expression derived there  was found  to be a reasonable fit to a well measured halo lag galaxy NGC~891 \citep{oos2007}. We discuss the form of the outflow/inflow factor $w(\zeta)$  in Sect.~\ref{sect:alphawind}, below.

\subsection{Time dependence}
\label{sect:time}
In this section, we use the logarithmic time, $T$, as defined in Eq.~\ref{eq:invariants}$\,$\footnote{When $\tilde\alpha_h$ is constant it is convenient to  multiply it into $\alpha$ in the log time equation  of Eqn.~\ref{eq:invariants} because this removes it from the equations to be solved.}.  When we recall that $R=r e^{-\delta T}$, we see from Eqns.~\ref{eq:field}, \ref{eq:A2b2}  and \ref{eq:invariants}, that 
\be
{\bf b}\propto e^{(2-a)\delta T}\propto (1+\alpha t)^{(2/a-1)}.\label{eq: t and f}
\ee
It is now clear that $1/\alpha$ (or $1/\tilde\alpha_h\alpha$ when $\tilde\alpha_h$ multiplies $\alpha$ in the definition of  $T$ in Eq.~\ref{eq:scaledT}) may be identified with a characteristic time (growth or decay) for the self-similar dynamo, but that $a$ determines the form of the growth function. 

This is where global conserved quantities  enter the scale invariant picture. If for example $a=1$, corresponding to a constant  global velocity or initial constant `seed' magnetic field, the field increases linearly in time at a rate given by $\alpha$ or $\tilde\alpha_h\alpha$.
If we suppose  instead that the the usual magnetic  helicity integrated over a volume $V$  is conserved, that is 
\be
\int~{\bf A}\cdot {\bf b}~ dV=constant,
\ee
then the global constant has Dimensions $L^6/T^2$. This requires $6\delta-2\alpha=0$ for invariance in time, and hence $a=\alpha/\delta =3$.
We conclude from Eqn.~\ref{eq: t and f} that the magnetic field must decline as $(1+\alpha t)^{-1/3}$. Such a decline agrees with previous suggestions
\citep[e.g.][]{Shuk2006,Black2015}.

Conserving the local magnetic helicity density \citep{M1978} itself, suggests $L^3/T^2$ as  the physical dimensions of a conserved quantity everywhere (gauge invariant over a closed volume however small), hence $a=3/2$ (a Keplerian value) from $3\delta-2\alpha=0$ and so the field grows slowly as $1+\alpha t)^{1/3}$.

If the  macroscopic `current' helicity ${\bf b}\cdot \nabla\times {\bf b}$ is conserved locally, then the local constant has Dimensions $L/T^2$ so that $a=1/2$, which gives rapid growth $\propto (1+\alpha t)^3$. The volume integrated version of this helicity  implies $L^4/T^2$ whence $a=2$ which implies a steady state. In our numerical studies we normally use $a=1$, which corresponds to either an invariant global velocity or  a constant initial `seed' magnetic field. Such a velocity may be set by a flat rotation curve, or possibly by a typical constant outflow velocity. An initial seed field must be determined by cosmology and a theory of galaxy formation  \citep[e.g.][]{WA2009,BeckAM2012,PMS2014}.

A limiting exponential growth in the dynamo field is given by supposing that there is a global constant with a fixed dimension of reciprocal time. This might be a fixed angular velocity $\Omega$ or possibly a constant rotational lag rate, but in any case the assumption requires $\alpha\rightarrow 0$ so that the relevant quantity is conserved in time.  In that case the limit of Eqn.~\ref{eq: t and f} as $\alpha\rightarrow 0$ gives  exponential growth as 
$b\propto e^{2\delta t}$. Here $\delta$ is a reciprocal  time characteristic of the  magnetized galaxy. If this is $r/v_\phi$ then the e-folding time is essentially a rotation period, as might be expected.

\section{Separate physical parts of the galactic magnetic dynamo}

In this section, we study the magnetic field produced when  either the sub scale turbulent dynamo ($\alpha_h$ term) or the convective flux freezing  in Eqns.~\ref{eq:dynamo1} is dominant. There are various examples in each case depending on the parameters. We also consider some important combinations of these effects. The objective is to isolate magnetic topology that may reveal which mechanism is important in a magnetized  edge on galaxy, Such data have been  reported in the recent CHANG-ES catalogue \citep{KIS2020}. Face on galaxies have already revealed their magnetic arm structure \citep[e.g.][]{Beck2015}.

{
  In order to compare the various effects, we will be referring to a parameter vector which is a subset of the form, $\{v1, w_o, w_1, \alpha_{h1}, K, r, \phi, z, C_1, C_2 \}$, where $r, \phi, z$ are coordinates.
  The exact description of these (scaled) parameters are found elsewhere in this paper.  However, for simplicity, it is helpful to keep in mind the following, with larger numbers representing a stronger effect:\\
  $v1$: rotational lag rate of the halo (Eq.~\ref{eq:lagonly})\\
  $w_0$: initial in-disc velocity of an outflowing wind (Eq.~\ref{eq:wonly})\\
  $w_1$: wind acceleration (Eq.~\ref{eq:wonly})\\
  $K$: a parameter describing the {\it similarity class} which is always $\equiv\,1$ in this paper (Eq.~\ref{eq:K})\\
  $C_1, C_2$: describes whether the solution is a dipole ($1,0$) or a quadrupole ($0,1$)\\
  $\alpha_{h1}$: a measure of the strength of the subscale turbulence (Eq.~\ref{eq:auxvariable})
  }

\subsection{Sub scale turbulent dynamo with halo rotational lag }
\label{sect:aux1}

The  auxilliary function $\tilde\alpha_h $ is first taken constant   and in this section, $w=0$.  Because of the radial dependence, the turbulent dynamo vanishes at small radius. This explicitly avoids any effect of an AGN.

It is convenient to write the logarithmic time $T$ according to 
\be
e^{\alpha T}\equiv (1+\tilde\alpha_h\alpha t),\label{eq:scaledT}
\ee
rather than the usual choice taken in Eqn.~\ref{eq:invariants}. 

This choice of logarithmic time removes $\tilde\alpha_h$ from the equation to be solved for the azimuthal vector potential, which becomes simply  
 (the prime indicates differentiation with respect to $\zeta$ )
\be
(1+\zeta^2)\tilde A_\Phi''+(K^2-v')\tilde A_\Phi=0.\label{eq:A3}
\ee
The constant $\tilde\alpha_h$ is constrained not to be strictly zero with this temporal scaling. Its value has an effect only on the time dependence of the field, once the reciprocal scale $\alpha$ ($a=1\,=>\,\alpha\,=\delta$) has been chosen for macroscopic reasons. We treat the pure flux conservation (sometimes referred to as `flux freezing') dynamo with $\tilde\alpha_d=0$ separately below. 

Subsequently the other components of the  vector potential follow as  (the middle expression restates Eqn.~\ref{eq:A3} when combined with the first and third)
\bea
K\tilde A_R &=&v\tilde A_\Phi-(1+v\zeta)\bar A_\Phi' \nonumber\\
K\tilde A_\Phi&=&\tilde A_R'+\zeta \tilde A_Z',\label{eq:A4}\\
K\tilde A_Z &=& \tilde A_\Phi+(v-\zeta)\tilde A_\Phi'.\nonumber
\eea
We have set 
\be
K\equiv 2-a.\label{eq:K}
\ee

The  scaled magnetic field can be shown to have the simple form
\bea
\tilde b_R&=&-\tilde A_\Phi',\label{eq:bR}\\
\tilde b_\Phi&=& \tilde A_R'+\zeta\tilde A_z',\label{eq:bPhi}\\
\tilde b_Z&=&\tilde A_\Phi-\zeta\tilde A_\Phi'.\label{eq:bZ}
\eea
These equations agree with those in \cite{HWI2018a} (the combined Eqns.~27 and 28 of that paper) when $\Delta=1$, $u=w=0$  and all diffusion terms are removed.

\subsection{Pure $\alpha_h$ Dynamo }
\label{sect:aux11}
If we set $v=0$  in the equations of Sect.~\ref{sect:aux1}, we obtain the pure $\alpha_h$ dynamo with radially increasing sub scale turbulent vorticity. Eqn.~\ref{eq:A3} becomes quite simple  and  is manifestly invariant under a change of sign of $\zeta$, although the derivative will change sign on crossing zero if $\bar A_\phi$ does not change sign. This occurs in what we call the dipolar solution below. The vector potential $\bar A_\phi$ can also change sign, which implies, by Eqn.~\ref{eq:bZ}, that  $\tilde b_Z$ passes through zero at the disc. This occurs for the quadrupole solution.
The equations for the magnetic field  (Eqns.~\ref{eq:bR}, \ref{eq:bPhi}, and \ref{eq:bZ}) show that consequently only the radial field component changes sign on crossing the equator for the dipole solution, but that both $b_Z$ and $b_\Phi$ change sign  for the quadrupole case. 
 
 The solution of Eqn.~\ref{eq:A3}  is  given in terms of hypergeometric functions (as described below for $v\ne 0$  in Sect.~\ref{sect:helicitypluslag}), although at small $\zeta$ there is a simple oscillatory limit.  In terms of the independent solution amplitudes $\{C_1,C_2\}$, the solution $\{1,0\}$ is the `dipolar solution' ($b_z$ continuous across the disc)  and $\{0,1\}$ is the `quadrupole solution' ($b_z$ zero at the disc). In the quadrupole case $
\bar A_\phi$ passes through zero and changes sign at the disc. Hence $b_R$ does not change sign on crossing the disc  but $b_Z$ and $b_\phi$ do. 

\begin{figure}{}
{\rotatebox{0}{\scalebox{0.9} 
{\includegraphics{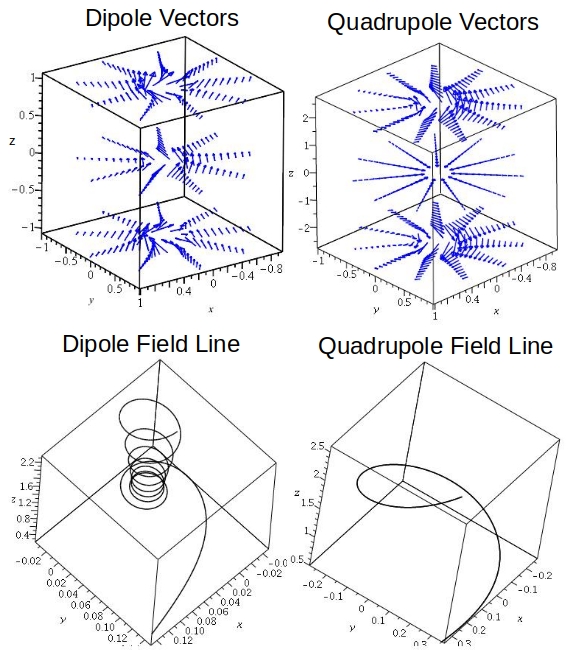}}}}
\caption{We show the dipole at left  and quadrupole at right in this figure, both above and below the plane. At upper left the local field direction is shown in 3d  for the dipole solution with only the sub scale turbulent helicity of Eqn.~\ref{eq:helicity} present. The similarity class $a=1$ ,and the length unit is the radius of the disc.  The spiral twist is very small except near the axis. The field line at lower left is for the dipole solution. It begins very near the axis at a large height above the disc and  descends while spiralling rapidly  to the point $\{r,\phi,z\}=\{0.22,0,0.5\}$.  Note that  the azimuthal field changes direction at large height. On the right the quadrupole solution shows less structure. The spiralling is very slow even at small radius (note the scale in the lower right hand  image)  and the  azimuthal field component does not change sign even at large heights. The quadrupole field line shown, starts at $\{r,\phi,z\}=\{0.5,0,0.5\}$ and rises. Both of these field lines are shown in clusters in figure (\ref{fig:fieldlinesalpha}).}
\label{fig:alphaeffect}
\end{figure}

\begin{figure}{}
{\rotatebox{0}{\scalebox{0.9} 
{\includegraphics{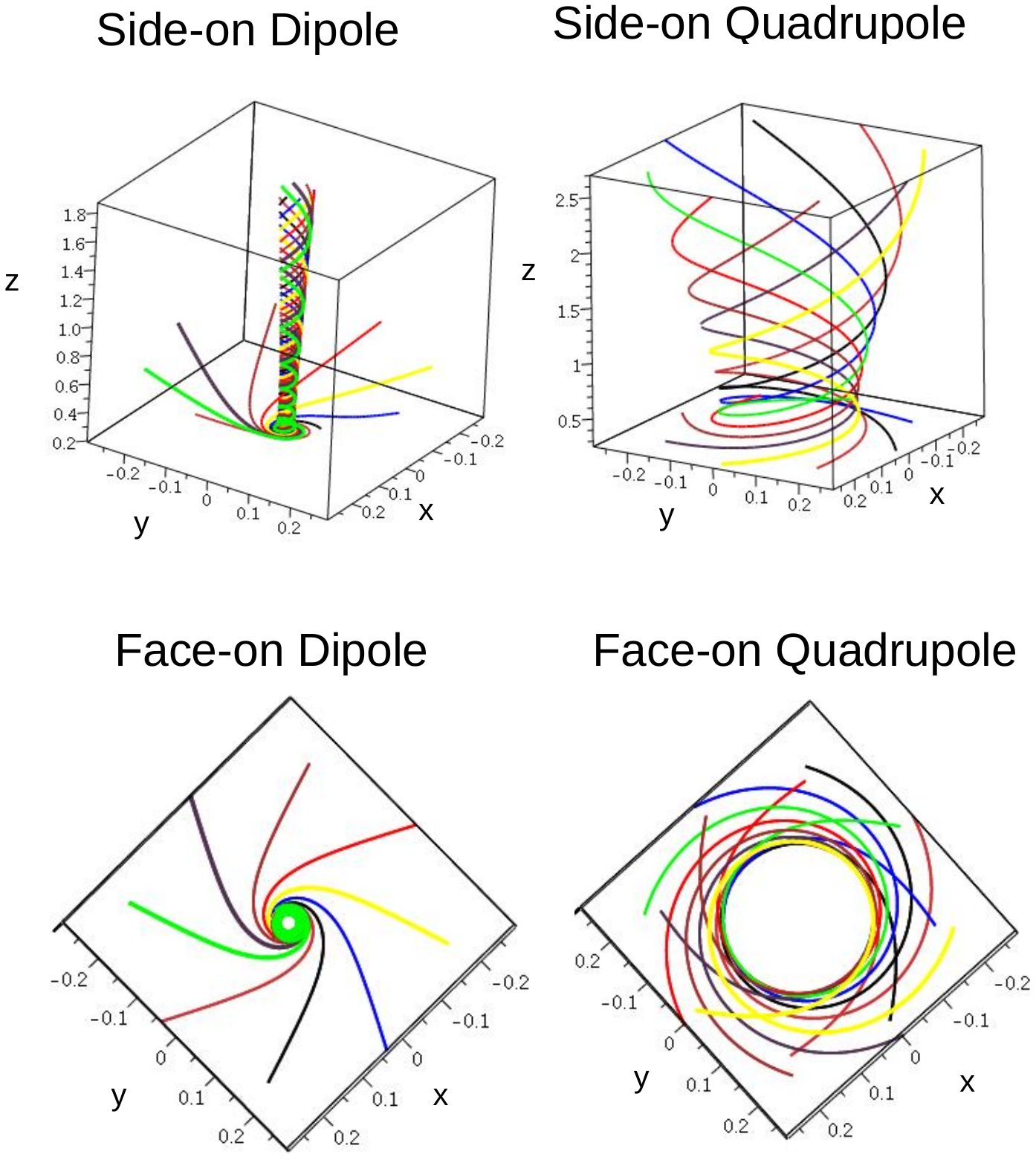}}}} 
\caption{The left column shows a collection of field lines for the dipole and the right column show the quadrupole.  The top row shows a side view and the bottom row shows a face-on view (the $\{r,\phi\}$ projection). 
  For the {\it dipole}, the lines begin on a circle of radius $r=0.22$ at height $z=0.5$ and they are spaced at $45^\circ$ around the circle. The field lines are descending on the central `funnel' and rising at large radius. One of these field lines (top left) is shown at lower left in figure (\ref{fig:alphaeffect}) for clarity. The rising  field lines indicate the connections between the outer  vectors at upper left in figure (\ref{fig:alphaeffect}).
  For the {\it quadrupole}, The field lines stand on radius $r=0.25$  at a height of  $0.25$. They are spaced at $45^\circ$ around the circle and are rising.
%
%
}
\label{fig:fieldlinesalpha}
\end{figure}

\begin{figure}{}
\rotatebox{0}{\scalebox{1.0} 
{\includegraphics{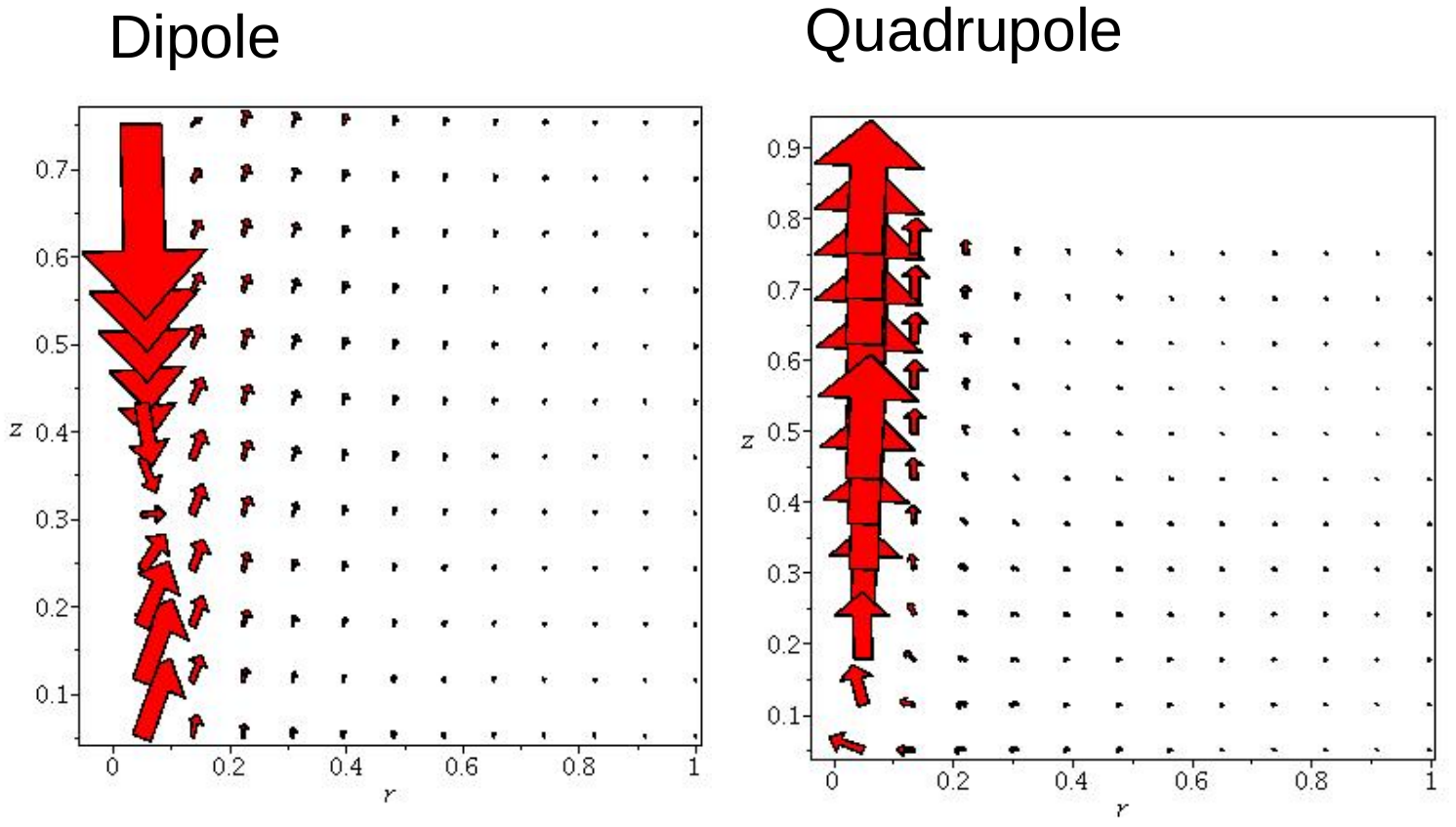}}}
\caption{The left figure shows the  pure $\alpha_h$ dynamo dipole in an $\{r,z\}$ cut.  The right figure shows the pure $\{\alpha_h\}$ quadrupole in an $\{r,z\}$ cut. The only parameter is $\tilde\alpha_h$ which appears only in the evolution time scale. It does not affect the field topology.   }
\label{fig:alphaeffectRZ}
\end{figure}

Conical symmetry is a characteristic of our version of self-similarity/scale invariance within a given galaxy.\footnote{This becomes a similarity symmetry between different galaxies which implies scale invariance in the family of galaxies when the length scale is the radius of the disc.} Hence we should not be surprised to find  evidence of this symmetry in each base solution of Eqn.~\ref{eq:dynamo1}.  Nevertheless the behaviour can be unexpectedly complex. We show some of the possibilities in Figures~\ref{fig:alphaeffect} and \ref{fig:fieldlinesalpha}, together with the projections in Figure~\ref{fig:alphaeffectRZ}. 


As a general comment on our figures, when field lines are shown they are accurate in 3d given the point at which we start them. In 3d vector plots, each vector shows the local direction of the magnetic field where the local point is either at the tail or the head of the vector.  When looking at 2d planar projections, the vectors are the component of the local magnetic field {\it in that plane}. Components perpendicular to that plane do not appear. Because the plotting routine plots the relative strength of vectors, these projections can give a false impression of the strength of the projected field. This must be taken into account when one tries  to reconcile the planar field with a 3d representation of the same field. The problem is not present when considering different views of the same 3d field lines.

Figure~\ref{fig:alphaeffect} demonstrates that, unless one is very near the galactic axis, the spirally rising field lines are very open. The dipole solution shows some interesting structure with height because near the axis the vertical and azimuthal field components  go through zero and reverse direction on the same side of the disc.  Normally referred to as a `parity change'. The field line at lower left is shown descending from  large height, but this is not necessary. The dipole field line 
ensemble in Figure~\ref{fig:fieldlinesalpha} begins at $z=1.8$. 
The quadrupole solution at upper right of  Figure~\ref{fig:alphaeffect} is  a slowly rising spiral. The single field line is assembled in a cluster spaced around a circle in Figure~\ref{fig:fieldlinesalpha}, the different lines corresponding to different colours. Only close to the axis does the quadrupole show `X field' behaviour in the $\{r,z\}$ plane. However  the dipole solution  shows a strong `X field'  throughout each quadrant. These poloidal projections are shown in Figure~\ref{fig:alphaeffectRZ} below.

Figure~\ref{fig:alphaeffectRZ} shows the $\{r, z\}$ cuts through the upper half of the  upper row images of  Figure~\ref{fig:alphaeffect}.  These figures correspond to the images in Figure~\ref{fig:fieldlinesalpha}. Only the  dipole shows a convincing high latitude `X field' in projection.  The axial  behaviour of the magnetic fields shows the field coming down  at $z>\sim 0.3$, and then rising at larger radii. At the same radius below $z\sim 0.3$, the field is rising as part of a field  that descended at smaller radii. 

The quadrupole image on the right shows magnetic field parallel to the disc plus an axial  rising (inside $r\sim 0.15$) magnetic field. We should recall Eqn.~\ref{eq:helicity} at this point, which indicates that the true sub scale turbulent vorticity goes as $\tilde\alpha_h\delta r e^{-\alpha T}$. When $K=a=1$, $\alpha=\delta$, and the quantity $\delta r$ is generally set so as  to give a macroscopic velocity in order of magnitude. That leaves $\tilde\alpha_h$ to determine the magnitude of the alpha effect.


One can only expect the pure $\alpha_h$ dynamo in galaxies for which there is little evidence of macroscopic jets or winds (unless the flow is  parallel  to the magnetic field). The macroscopic ($\alpha/\Omega$) dynamo in our model occurs  either due to the lagging rotation in the halo  or because of a gradient in the outflow/inflow velocities.  There are essentially no parameters in the magnetic field of this section beyond the choice of the self similar class, $a$, which we have set equal to $1$ , reflecting a constant global velocity. This encourages us to search for this coherent magnetic field behaviour in the observations.  One might expect this behaviour to be correlated with vigorous star formation.

\begin{figure}{}
\begin{tabular}{l} 
\rotatebox{0}{\scalebox{0.8} 
  {\includegraphics{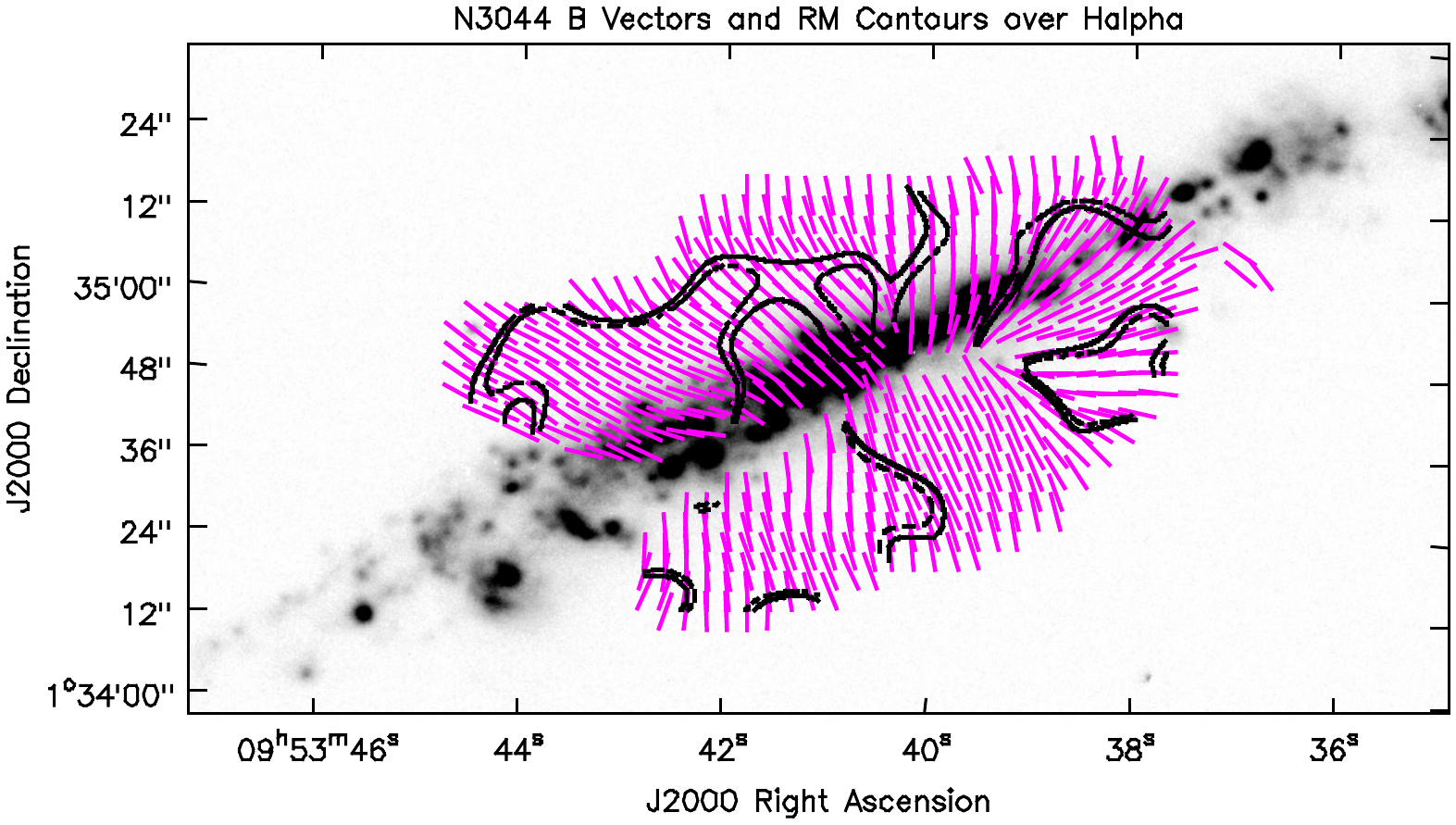}}}\\
\vspace{0truein}\\
\rotatebox{0}{\scalebox{0.8} 
  {\includegraphics{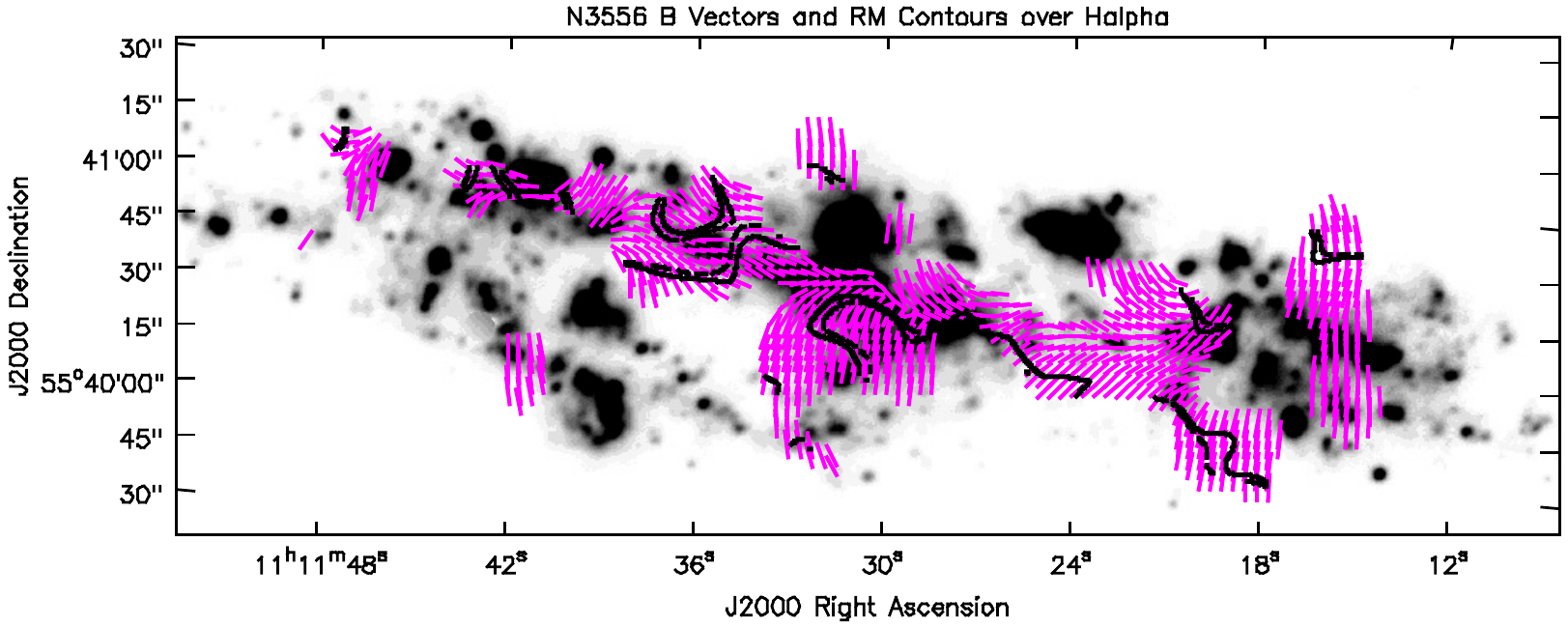}\hfill}}
\end{tabular}
\caption{NGC~3044 (top) and NGC~3556 (bottom) with H$\alpha$ emission shown in greyscale. 
B vectors, corrected for Faraday rotation, are shown in magenta. RM contours have been set only for the purpose of indicating where there is a sign change from positive to negative or vice versa.
The contours are set to +10 (solid curves) and -10 (dashed curves) rad m$^{-2}$ for NGC~3044 and at +15 (solid curves) and -15 (dashed curves) rad m$^{-2}$ for NGC~3556 to help guide the eye. That is, the sign change occurs in the narrow gap between the solid and dashed contours. 
See \protect\cite{KIS2020} for more explanation of the polarization data and \protect\cite{varg18} for information on the H$\alpha$ image. 
All data have been downloaded from queensu.ca/changes. }
\label{fig:alphaHalpha}
\end{figure}

Referring to the observed fields displayed in the CHANG-ES catalogue of \cite{KIS2020}, NGC~3044 may be a candidate for a pure star formation (i.e. sub scale turbulence resulting in $\alpha_h$) dynamo (see top of Figure~\ref{fig:alphaHalpha} and the figures of this section). The base function would be predominantly dipole. The characteristic magnetic behaviour is observed to be predominantly high-latitude `X field' with no parallel equatorial component. Moreover the field falls off rapidly radially but extends far into the halo. The very strong axial field is a formal feature of Eqn.~\ref{eq:dynamo1} when $\alpha_h\rightarrow 0$ as $r\rightarrow 0$, because the spatial gradient  times $\alpha_h$ is required to match the time derivative. There is also the $1/r$ field dependence that reflects the self-similar global symmetry.
Figures~\ref{fig:alphaeffect}, \ref{fig:fieldlinesalpha} and \ref{fig:alphaeffectRZ} all illustrate these properties for the $\alpha_h$ dipole.

The rotation measure (RM) given in \cite{KIS2020}  and displayed in rudimentary form by contours in Figure~\ref{fig:alphaHalpha} varies as might be expected from rising, spiralling, field lines.
In between the double-lines in the galaxy figures (solid and dashed curves close to each other) is where the RM sign changes. On the solid curve side, the RM is positive (field lines pointing towards us) and on the dashed curve side, the RM is negative (field lines pointing away from us).
The RM also shows a sign change across the disc in some regions which can correspond to a change in sign of radial field.  Figures~\ref{fig:alphawind3} and \ref{fig:analyticFF}  below show that the $\alpha_h$ dynamo with wind or even pure flux freezing with wind may also fit the data. NGC~3448 (not shown) may be a more depolarized second example. 

NGC~3556 is another extreme where a coherent galactic field is scarcely established. It is likely that it is dominated by a series of local dynamos driven by local star formation, so we are actually seeing the turbulent nature of sub scale vorticity.  That is, there are multiple $\alpha_h$ dynamos acting.  The data are shown for comparison in the bottom image of Figure~\ref{fig:alphaHalpha}.

\newpage

\subsection {Halo rotational lag and the $\alpha_h$ dynamo}
\label{sect:helicitypluslag}


The general solution to Eqn.~\ref{eq:A3}  with $v'=-v_1$ for $\tilde A_\Phi(\zeta)$ becomes 
\be
\tilde A_\Phi(\zeta)=C1~(1+\zeta^2)F(\alpha_1,\beta_1,\gamma_1,-\zeta^2)+C2~\zeta(1+\zeta^2) F(\alpha_2,\beta_2,\gamma_2,-\zeta^2),\label{eq:Asol}
\ee
where $F(\alpha,\beta,\gamma,z)$ is the hypergeometric function. The arguments are (with $n=1,2$)
\bea
\alpha_n&=&(\frac{3}{4}+\frac{n-1}{2})+\frac{\sqrt{1-4(K^2+v_1)}}{4},\nonumber\\
\beta_n&=&(\frac{3}{4}+\frac{n-1}{2})-\frac{\sqrt{1-4(K^2+v_1)}}{4},\label{eq:arguments}\\
\gamma_n&=&\frac{1}{2}+(n-1).\nonumber
\eea
In our examples  the solution $\{C1,C2\}=\{1,0\}$ is 'dipolar' in form with $b_Z$ non-zero at the disc, while the solution $\{C1,C2\}=\{0,1\}$ is quadrupolar  and has zero $b_Z$ at the disc. 

From Eqn.~\ref{eq:helicity}, we note that when $\tilde\alpha_h$ is constant we may use its value  to define the sub scale turbulent velocity $ <1$ and so let the velocity scale $\delta r$  be macroscopic. That is, $\delta r$ reflects mainly lagging rotational velocity $v$ or wind velocity $w$ (although we do not use a `wind' velocity in this section). Consequently Eqn.~\ref{eq:velocity} shows that the magnitude of $v$ and $w$ should be chosen $O(1/\tilde\alpha_h)$  so as to reflect the larger macroscopic rotation and wind magnitudes. It is likely that $\tilde\alpha_h=O(0.1)$ \citep[e.g.][]{Tam2009} so that $v_1=O(1)$ \citep{H2007}\footnote{The halo rotational lag rate was found to vary from  $~40$ km/sec/kpc to $~15$ km/sec/kpc in \cite{oos2007}.} and $w_o$ and $w_1$ may be $O(10)$ in extreme cases \citep[e.g.][]{He2018}. Extensive supernova activity may lead to larger sub scale velocities locally.




\begin{figure}{}
\rotatebox{0}{\scalebox{1.0} 
{\includegraphics{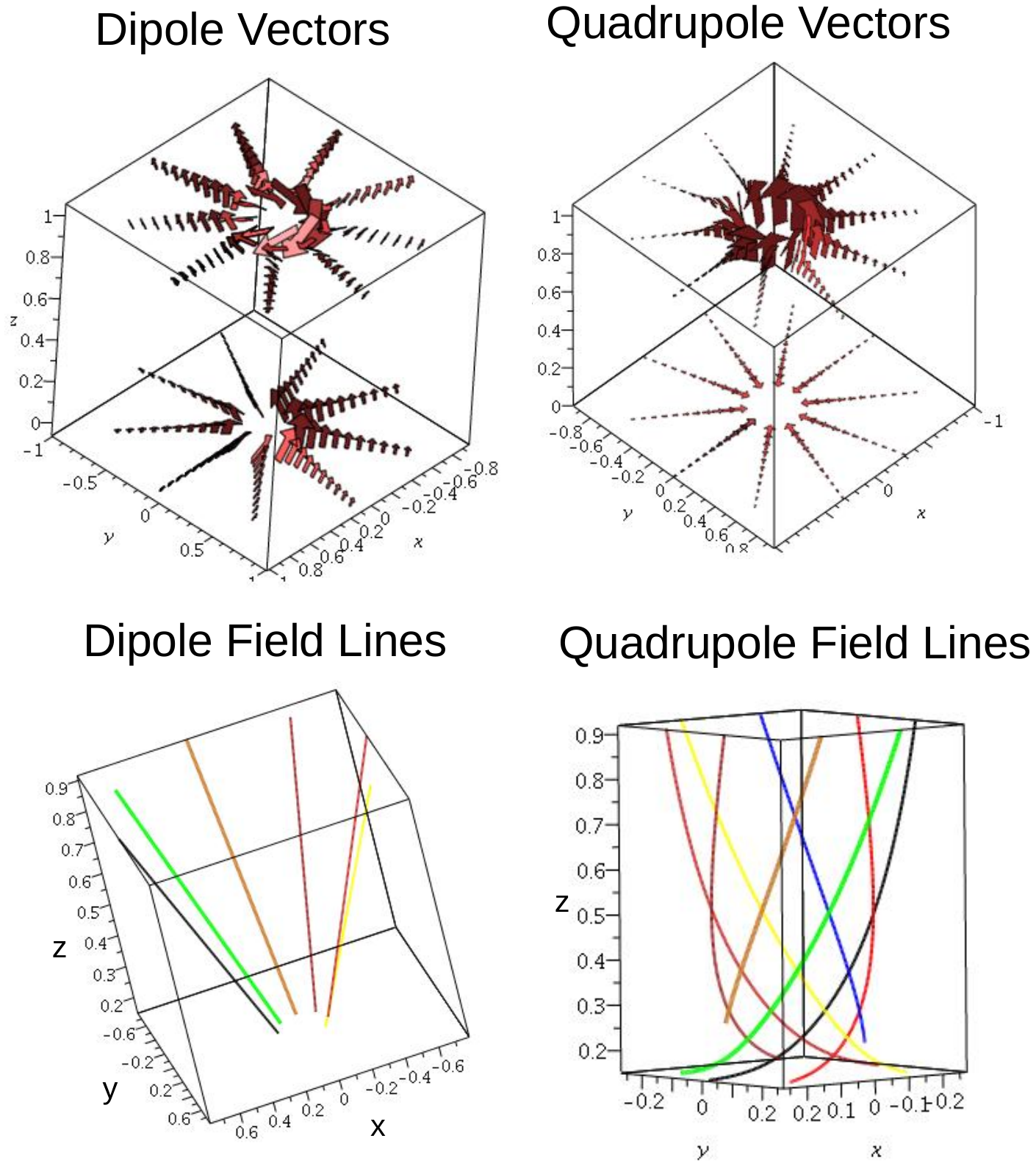}}}
\caption{
The top row shows the  magnetic vectors in 3d for the dipole on the left and the quadrupole on the right. The bottom row shows clusters of field lines for the dipole on the left and the quadrupole on the right.
  The field lines of the {\it lower row} start at $z=0.15$, $r=0.25$,  and then in steps of $\pi/4$ around the circle indicated by different colours. Only half the circle is shown for clarity. The rotational lag rate is $v_1=1$,
  $K=a=1$ for all of these examples.
  For the magnetic field vectors of the {\it top row}, only two levels in $z$ are shown for clarity, but they may be connected by referring to the 3d images of the second row.
 }
\label{fig:alphalag1}
\end{figure}

In Figure~\ref{fig:alphalag1}, we show a dipole solution and a quadrupole solution in 3d on the top row and  a cluster of corresponding field lines for each case in the bottom row.  We show only the upper half plane for clarity, but $b_\Phi$ and $b_Z$ change sign on crossing the disc in the quadrupole solution, while only $b_R$ does so in the dipole solution. 

\begin{figure}{}
\rotatebox{0}{\scalebox{1.0} 
{\includegraphics{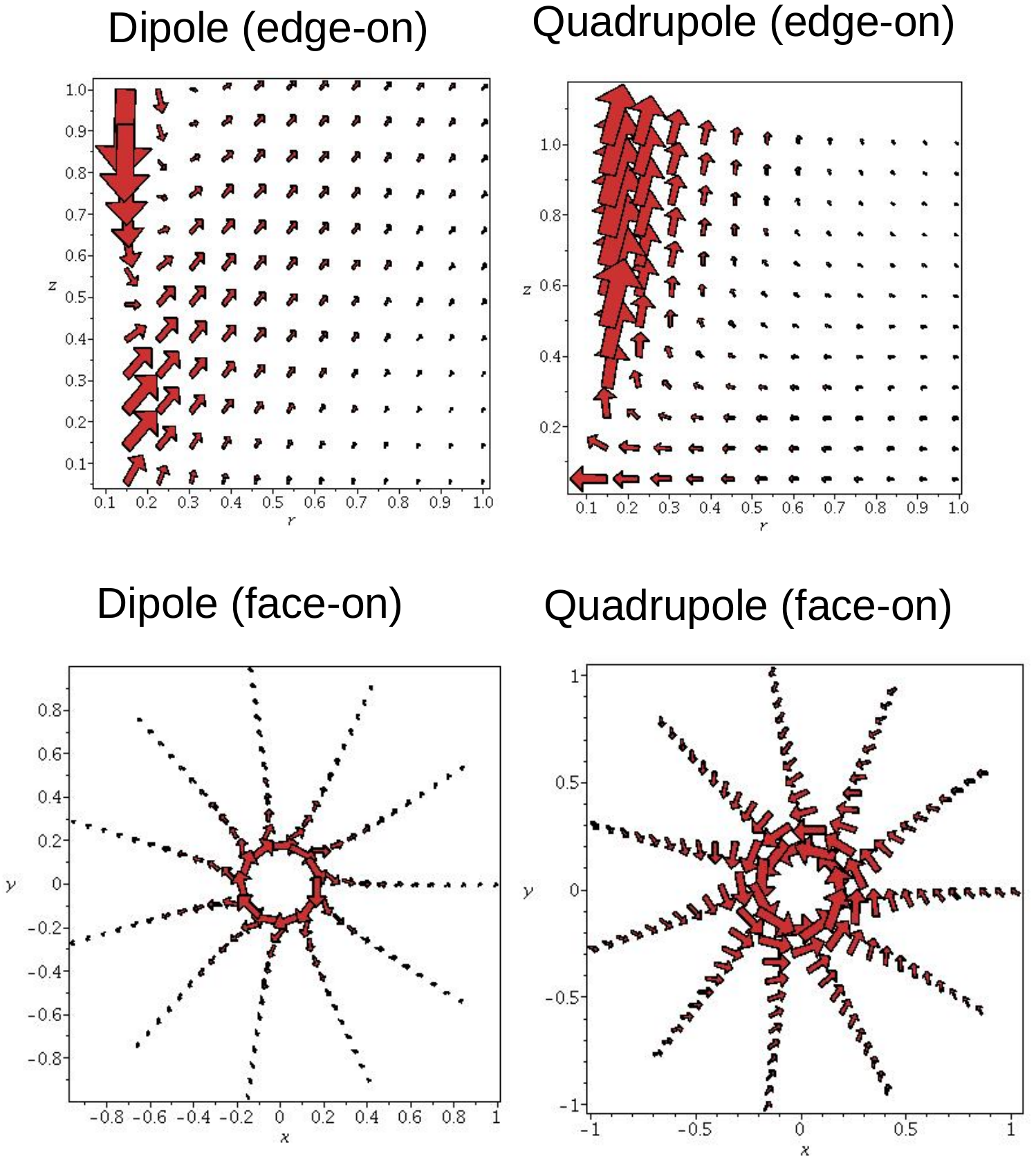}}}
\caption{
  In the upper row, the left column shows the dipole magnetic field for the case
  $\{v_1,K,C1,C2\}=\{1,1,1,0\}$ projected  into the $\{r,z\}$ poloidal plane, integrated from $r=015$ and $z=0.05$.  This contrasts the X field behaviour.
  The right column shows the same projection for the quadrupole solution with parameters  $\{1,1,0,1\}$.
  The lower row shows the corresponding  projections in the toroidal plane at $z=0.5$ }
\label{fig:alphalag2}
\end{figure}

We see that the quadrupole solution produces a projected `X type' geometry (recall the axial  symmetry)  at  small polar angle  that includes the axis, while the dipole solution produces the X geometry at much larger polar angles. The dipole field lines appear to be straight in this representation because the poloidal field is much larger than the toroidal field. However they do turn in azimuth as is shown in Figure~\ref{fig:alphalag2}.  


The dipole has an interesting structure at large height  where the field lines become  tightly wound near the axis. The winding sense is opposite to the sense of the exterior winding, which implies that the azimuthal field passes through zero at some height and radius. This  leads to a sign change in the  RM with latitude that can be seen in an RM plot \citep{HWI2018a}, but we do not show this here for brevity.  

The addition of rotational halo lag has not  greatly changed the topology from that of the pure $\alpha_h$ dynamo as seen in Figures~\ref{fig:fieldlinesalpha} and \ref{fig:alphaeffectRZ}. The dipole solution changes the rotation sense at height near the axis but remains similar at large radius, showing a strong X field. The quadrupole field is  less tightly wound in the presence of halo lag and shows `X field' only in a very limited range of latitudes. At this point, X field magnetic topology favours very much a dipole symmetry across the galactic equator.




 
Figure~\ref{fig:alphalag2} shows poloidal and toroidal cuts of the same fields shown in Figure~\ref{fig:alphalag1}.  The quadrupole solution  shows  very similar behaviour to the pure alpha solution although it is less tightly wound. The dipole solution has the very tight winding near the galactic axis as does the pure turbulent field and has a very strong X field behaviour generally.  The sign of the pitch angle of the dipole magnetic field changes sign with height and radius. Such behaviour is more in accord with the average CHANG-ES behaviour, as shown in \cite{KIS2020}.

{\it In terms of the observations reported in \cite{KIS2020}, Figures~\ref{fig:alphalag2}  and \ref{fig:alphaeffectRZ} show that it would be difficult to detect the effect of moderate halo rotational lag on the $\alpha_h$ dynamo for  the dipole, except near the axis. However  near  the axis tight winding can lead to beam depolarization as for the dipole in Figure~\ref{fig:alphalag1}.  This often seems to be the case observationally. The quadrupole does not deliver a convincing X field.}

We turn in the next section to a similar study of the magnetic effects of a combined accelerating wind  and halo rotational lag.

\subsection{Accelerating disc wind plus  halo rotational lag acting on the turbulent dynamo}
\label{sect:alphawind}

Given  the (sub scale) turbulent dynamo with constant $\tilde\alpha_h$ plus  either outflow or accretion and rotational halo lag, Eqns.~\ref{eq:dynamo1} plus scale invariance  lead to the equations for the scaled vector potential as
\bea
K\tilde A_R&=& -\tilde A_\Phi'(1+\zeta v(\zeta))+v(\zeta)\tilde A_\Phi-w(\zeta)(\tilde A_R'+\zeta \tilde A_Z'),\nonumber\\
K\tilde A_\Phi&=& \tilde A_R'+\zeta \tilde A_Z'-w(\zeta)\tilde A_\Phi',\label{eq:vecpotw}\\
K\tilde A_Z&=& \tilde A_\Phi+(v(\zeta)-\zeta) \tilde A_\Phi'.\nonumber
\eea
These may be combined to give the equation for $\tilde A_\Phi$ as
\be
(1+w(\zeta)^2+\zeta^2)\tilde A_\Phi''+2w(\zeta)(K+w(\zeta)')\tilde A_\Phi'+(K(K+w(\zeta)')-v(\zeta)')\tilde A_\Phi=0.\label{eq:vecpotPhiw}
\ee
The scaled magnetic field components are given as in Sect.~\ref{sect:aux1}. These can be expressed  solely in terms of the azimuthal vector potential and its derivative plus the auxiliary  velocities, by using Eqns.~\ref{eq:vecpotw}. 

The forms for the auxiliary  scaled vertical velocity and the scaled rotational lag (the physical velocity is given in Eqn.~\ref{eq:velocity}) are taken as 
\bea
w(\zeta)&=&w_o+w_1\zeta,\label{eq:wonly}\\
v(\zeta)&=&-v_1\zeta,\label{eq:lagonly}
\eea
where $w_o$ gives the scaled vertical velocity at the disc, $w_1$ is the scaled rate of acceleration with $\zeta$ and  $v_1$ gives the scaled halo lag. If $w_o$ and $w_1$ are negative then we have an decelerating flow onto the disc.  If $w_1>0$  with negative $w_o$ then we may have an accelerating inflow. 
The rotational lag rate $v_1$ is generally positive. One should note that near the axis of the galaxy $\zeta\rightarrow\infty$.  Either scaled velocity  is then singular above the nucleus, but the physical velocity (\ref{eq:velocity}) simply increases with height. 

\begin{figure}{}
\rotatebox{0}{\scalebox{1.0} 
  {\includegraphics{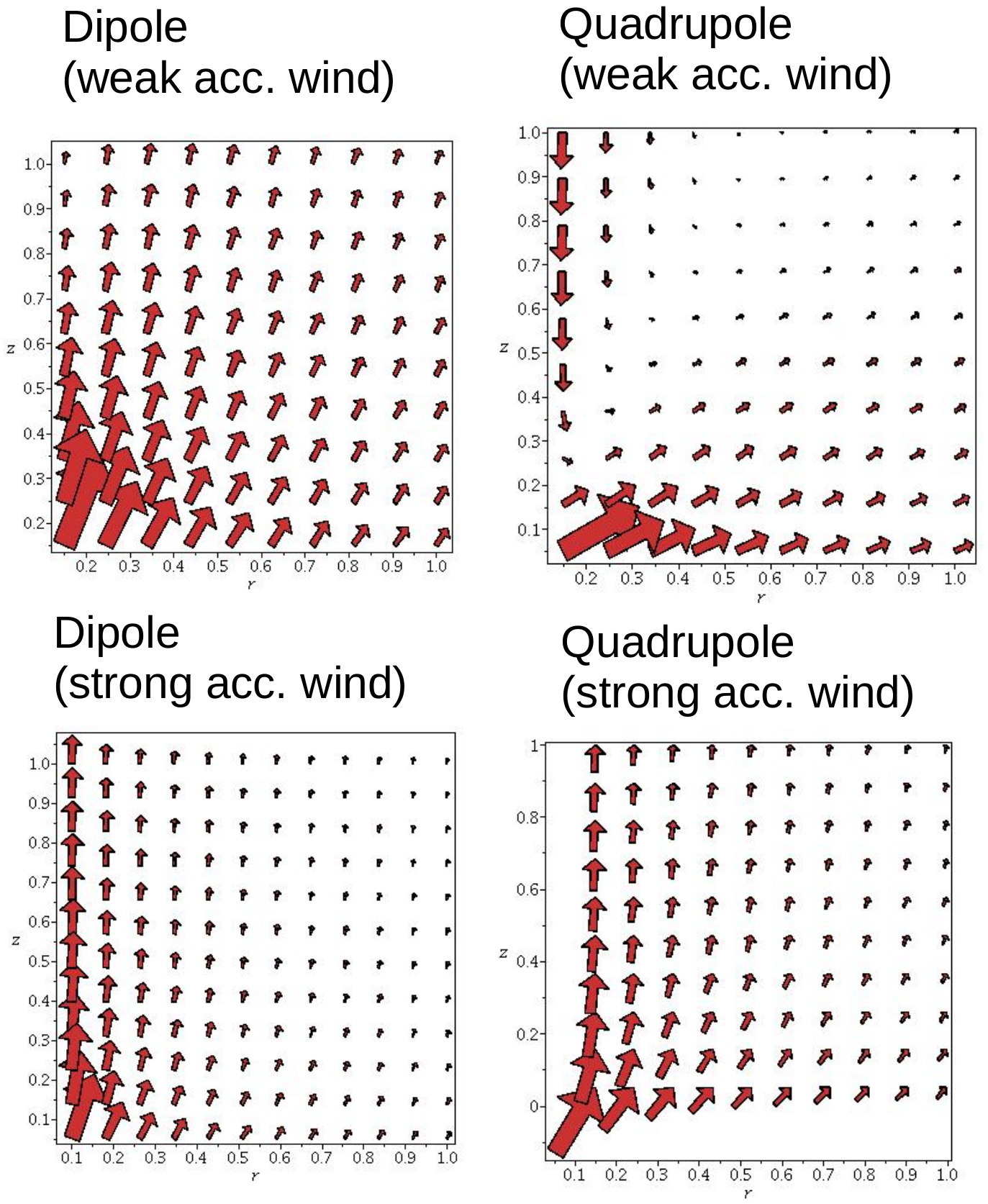}}}
\caption{Here, the magnetic field is subject to a disc wind and vertical acceleration in an outflow in addition to sub scale turbulence. The {\it left hand column} shows the dipole field  in the $\{r,z\}$ plane with parameter vector $\{v1,w_o,w_1,K,r,z,C_1,C_2\}$ equal to $\{0,1,1,1,r,z,1,0\}$ at upper left (a weaker wind) and $\{0,1,3,1,r,z,1,0\}$ at lower left (a strong wind).  
  The strong accelerating wind at lower left has pulled the field nearly parallel to the axis.
 The {\it right column} shows the quadrupole solution with parameter vector $\{0,1,1,1,r,z,0,1\}$ at upper right and $\{0,1,3,1,r,z,0,1\}$ at lower right. At lower right the magnetic field is rendered almost entirely vertical by a gradient in the vertical wind velocity. }
\label{fig:alphawind1}
\end{figure}

\begin{figure}{}
\rotatebox{0}{\scalebox{1.0} 
{\includegraphics{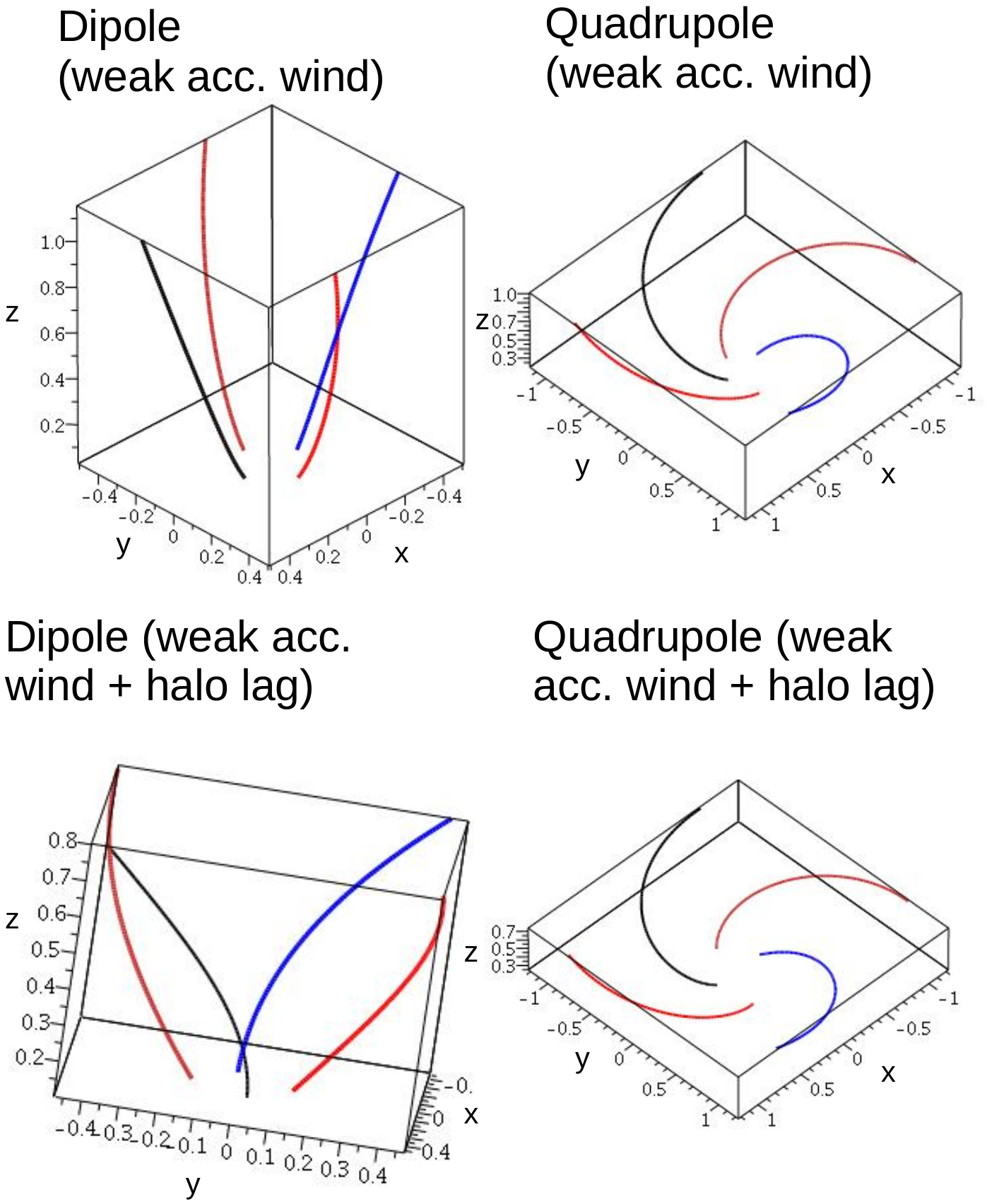}}}
\caption{ All the field lines in different panels of this figure are  started at $r=z=0.25$ and $\phi=0$.  At {\it upper left}, a cluster of field lines for a dipole case is shown taken from the image at upper left in figure(\ref{fig:alphawind1}). The parameter set is $\{v1,w_o,w_1, K, r,\phi,z,C_1,C_2\}$ = $\{0,1,1,1,r,\phi,z,1,0\}$. The {\it lower left} panel is for a dipole with parameter set $\{1,1,1,1,r,\phi,z,1,0\}$. The field lines are placed at cardinal points around the initial circle and height.
  At {\it upper right}, four quadrupole field lines that are taken from the  $\{r,x\}$ image at upper right in figure (\ref{fig:alphawind1}) are shown.  The parameter list is $\{0,1,1,1,r,\phi,z,0,1\}$ for the quadrupole at upper right.
  The lower right cluster is for a quadrupole magnetic field with the parameter set $\{1,1,1,1,r,\phi,z,0,1\}$.
}
\label{fig:alphawind2}
\end{figure}

Eqn.~\ref{eq:vecpotPhiw} may now be solved in terms of hypergeometric functions  with complicated arguments, according to MAPLE 2019. We proceed to express a few results for `typical' parameters. In Figures~\ref{fig:alphawind1} and \ref{fig:alphawind2}.

In Figure~\ref{fig:alphawind1}, we show sample `dipole' and  `quadrupole' magnetic fields  with and without a strong wind  acceleration.  The cuts are in the $\{r,z\}$ plane and since the field is axially symmetric the $3d$ topology is easily inferred. The macroscopic motions act on the sub scale magnetic field that is present due to turbulence ($\alpha_h$ dynamo). The  constant auxiliary function $\tilde\alpha_h$ does not appear in the equations except in the  temporal evolutionary Eqn.~\ref{eq:scaledT}. The  magnetic topology is very sensitive  to an accelerating wind velocity. Such a gradient generates an $\alpha_h/\Omega$ (macroscopic) dynamo  just as a halo rotational lag or a gradient of angular velocity in the disc. 

It is not shown here but we find that a strong  disc wind ($w_o\le\sim 10$) without strong acceleration only convects the boundary conditions at the disc to large heights, with little change in the topology. For this reason we have focussed on the acceleration in the examples.


 What this figure emphasizes over all, is the strong dependence on the wind gradient of the magnetic topology in the halo with a moderate initial disc wind. The `quadrupole' solution  shows `X field' structure only with the wind acceleration. The dipole field gives a strong `X field', particularly {\it without} wind acceleration. 

By varying the initial disc wind $w_o$ and the acceleration rate $w_1$, intermediate cases can be found. The field is also combed out by accelerating accretion because there is no barrier at the disc in the model. Decelerating accretion can be arranged with both $w_o$ and $w_1$ negative.

The quadrupole column in Figure~\ref{fig:alphawind2} compares a moderate accelerating wind with no halo rotational lag (upper panel) to one with a halo rotational lag, comparable to the accelerating wind velocity at the disc. The flatter spiral generation with the presence of halo lag is quite marked, because the lines in the lower panel rise only to a height of $0.7$ rather than $\sim 1$. We recall that the unit of length is the radius of the galactic disc.

A similar flattening effect is observed with the same change in rotational  halo lag for the dipole, as  shown in the left hand column panels. Because it appears  (Figure \ref{fig:alphawind1}) that the dipole is the principal source of strong X fields, we can recognize the effects of halo lag in the increasing polar angle (decreasing halo latitude) of the magnetic polarization with increasing lag rate.



\begin{figure}{}
  \begin{tabular}{c}
\vspace{-0.1truein}\\
\rotatebox{0}{\scalebox{0.9} 
  {\includegraphics{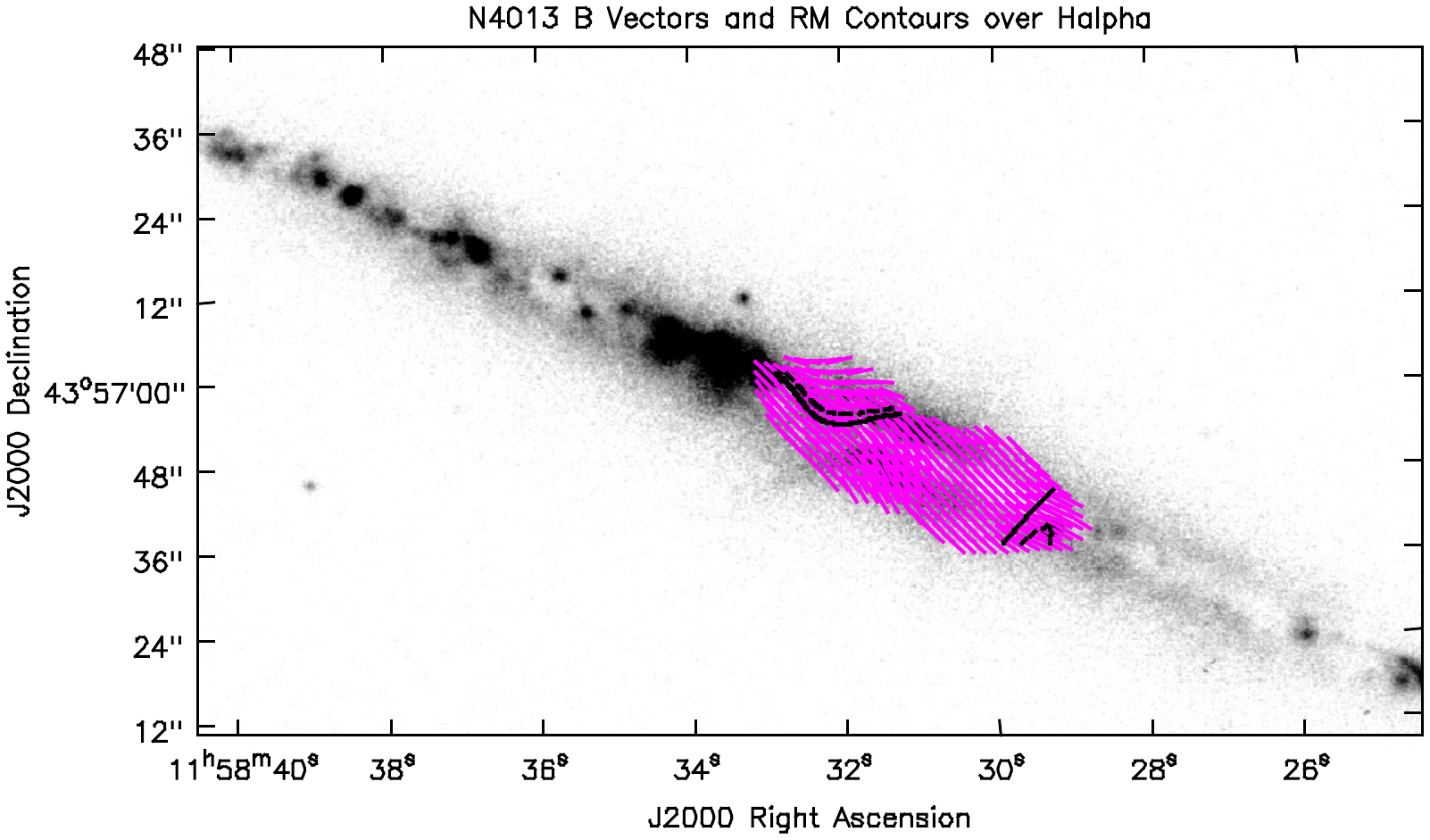}}}
    \end{tabular}
\caption{ The  image shows NGC~4013 with H$\alpha$ emission shown in greyscale. 
B vectors, corrected for Faraday rotation, are shown in magenta. RM contours have been set only for the purpose of indicating where there is a sign change from positive to negative or vice versa, and are present to help guide the eye.
The contours are set to +10 (solid curves) and -10 (dashed curves) rad m$^{-2}$ 
See \protect\cite{KIS2020} for more explanation of the polarization data and \protect\cite{varg18} for information on the H$\alpha$ image. 
All data have been downloaded from queensu.ca/changes. }
\label{fig:alphawind3}
\end{figure}

 Adding a halo rotational lag adds to the complexity of the magnetic topology. In particular it can lead most noticeably to a reversal of the magnetic helicity with height, which leads to the observed reversal in `parity' in the RM. This is similar to the behaviour found in \cite{HI2016}. The lag also increases the extent of the projected `X field' near the disc.

 Figure~\ref{fig:alphawind3} shows the CHANG-ES polarization (with Faraday Depth sign change) data for the galaxy NGC~4013. The X magnetic field is wound very tightly and is really only visible very close to the disc. This might suggest, according to Figures~\ref{fig:alphawind1} and \ref{fig:alphawind2}, that the coherent magnetic field in NGC~4013 is  a quadrupole with considerable halo rotational lag.  However, close inspection of  Figure~\ref{fig:alphawind3} reveals that the field is more like an inverted V than an X. That is, the field converges on the axis above the disc. There is weak indication of this  for the quadrupole in Figure~\ref{fig:alphalag2}.

We conclude from this section that a vigorous constant wind can carry equatorial boundary conditions to large heights. However a gradient in the wind velocity is more important for the magnetic complexity such as strong deviations from a pure $\alpha_h$ dynamo.   The main effect of a strong wind is to transport magnetic field to greater height and to remove the `X field' in favour of a  poloidal field perpendicular to the galactic disc. The behaviour is particularly pronounced near the axis. The rotational halo lag along with the wind acceleration, acts as an $\alpha/\Omega$ dynamo given the $\alpha_h$ sub scale turbulent average dynamo. The rotational lag gives a field strongly wound near the disc while the wind acceleration stretches the field vertically.

Reviewing the observations in the CHANG-ES catalog, we note  that NGC~3044  (top of Figure~\ref{fig:alphaHalpha}) may be compatible with a moderate disc wind and acceleration as seen for the dipole in Figure~\ref{fig:alphawind1} at upper right. This is more consistent with the extended halo in NGC~3044 than is the pure $\alpha_h$ dynamo of Sect.~\ref{sect:aux11}.

 Quadrupole behaviour seems rarer but may be illustrated in NGC~4013 (Figure~\ref{fig:alphawind3}). We refer to both the CHANG-ES catalog \citep{KIS2020} and Figure~12 in \cite{SDMI}. At upper right in Figure~\ref{fig:alphawind1}, we see a quadrupole with field largely parallel to the disc  (i.e. a very low latitude X field) that falls off rapidly into the halo. The model predicts therefore large angle but weak X field, which corresponds to Figure~12 in \cite{SDMI}.  The data from \cite{KIS2020} is shown in  Figure~\ref{fig:alphawind3}.  The sign change in the RM across the disc in NGC~4013 can also be due to a quadrupole.

\subsection{A pure macroscopic `dynamo'}
\label{sect:aux2}

We consider the special case when there is no  active  $\alpha_h$ effect  combined with macroscopic motion, as there has been in all previous cases. It should be described nevertheless  as an $\alpha/\Omega$ dynamo because there must be an initial field for the velocity to act on. With $a=1$ and $K=1$, flux {\it density} is  conserved  (equivalent  Dimensionally to a magnetic field or velocity constant) and a  local field grows linearly according to Sect.~\ref{sect:time}. 

The existence of a global magnetic flux constant requires $a=3$ and hence $K=-1$.  Global flux conservation can only produce topological enhancements in the magnetic field which will decline  slowly in time.  We recall the discussion of time dependence in Sect.~\ref{sect:time}, which shows the magnetic field to decrease in time as $(1+\alpha t)^{-1/3}$ with $a=3$. 

We set $\tilde\alpha_h=0$ in this section and so Eqn.~\ref{eq:invariants} defines the logarithmic time. Because we also suppress the diffusion in this paper, we are left with the pure flux freezing  during the macroscopic motion  as described by  the second term in Eqn.~\ref{eq:dynamo1}.    We continue with our scale invariant formulation, which prescribes the velocity field according to Eqn.~\ref{eq:velocity} and the magnetic field according to Eqns.~\ref{eq:field} and \ref{eq:A2b2}. We proceed with only rotational and outflow/inflow velocities so that  the radial velocity $u=0$. This limit is related to the study performed in \cite{HI2016} using dynamical equipartition and a Mestel disc, but here the assumption of scale invariance dictates the velocity field, rather than  a dynamical model.

The reason for studying this limiting case is due to the predictive nature of the analytic magnetic fields with regard to scale heights of the magnetic field. 
This is related to the recent results for CHANG-ES galaxies reported in \cite{KIW2018}.

The surviving term in Eqn.~\ref{eq:dynamo1} requires solving a simple set of equations for the scale invariant vector potential namely
\bea
K\tilde A_R&=&v(\tilde A_\Phi-\zeta \tilde A_\Phi')-w(\tilde A_R'+\zeta\tilde A_z'),\nonumber\\
K\tilde A_\Phi&=&-w\tilde A_\Phi',\label{eq:FFvecpot}\\
K\tilde A_Z&=& v\tilde A_\Phi'.\nonumber
\eea
 Note that it is the outflow/accretion that drives the whole system. Setting $w=0$ renders the system trivial.
  
Eqns.~\ref{eq:FFvecpot} are easily solved analytically when we take the halo rotational lag  in the pattern frame and the outflow velocity  in the familiar forms
\bea
v&=&-v_1\zeta,\nonumber\\
w&=&w_o+w_1\zeta,\label{eq:FFvelocity}
\eea
However even the  case, wherein $v$ and $w$ are both constant (so that  physical velocities are proportional to $r$ from Eqn.~\ref{eq:velocity}, is  of  interest as it is both analytic and simple. The field components  follow ultimately  from Eqns.~\ref{eq:bR} to \ref{eq:bZ}. These imply with $v$ and $w$ constant and Eqns.~\ref{eq:FFvecpot},  that  
\bea
b_R&=&-\frac{e^{(2\delta-\alpha)T}}{rw}KC_1e^{-\frac{K}{w}\frac{z}{r}},\nonumber\\
b_\Phi&=& \frac{e^{(2\delta-\alpha)T}}{rw}e^{-\frac{K}{w}\frac{z}{r}}(C_1v-C_2K),\label{eq:FFcstvel}\\
b_Z&=& \frac{e^{(2\delta-\alpha)T}}{r}C_1e^{-\frac{K}{w}\frac{z}{r}}(1+\frac{K}{w}\zeta).\nonumber
\eea

From these relations we observe that the exponential decline of the magnetic field in $z$ at fixed $r$, is slower with larger wind velocity, so that the scale height increases.  The  exponential decline is also slower at larger $r$ for a constant outflow, which gives the magnetic (or radio) halo a `butterfly' shape. The azimuthal field increases with increasing $v$ and decreases with increasing  $w$ at the disc. The radial field also decreases with increasing $w$ at the disc. Boundary conditions at the disc are `lifted' into the halo. This behaviour has also been found in the previous section.

 In general, an individual field line always extends in the same sense in radius, azimuth, and height so that it appears as a rising, expanding helix, much as shown in \cite{HI2016}.  It differs from the pure $\alpha_h$ dynamo of the  Sect.~\ref{sect:aux11}, in that the spirals are detectable at much larger radii.  This topology already gives an X type field  near the disc  when projected into the $\{r,z\}$ plane.

 We note  from the solution that  $b_R$  changes sign with $w$ on crossing the disc. However there is no halo rotational lag in $z$ with $v$ constant, but there is in radius by Eqn.~\ref{eq:velocity}. Therefore there is no need for $v$ to change sign on crossing the disc.  Moreover $b_\Phi$  also changes sign on crossing the disc. The component $b_Z$ does not change sign  because $\zeta$  also changes sign. Thus the field obeys a dipole symmetry, but with a change in sign of both tangential field components. 
 
 As a result of these symmetries, there is a change in the magnetic helicity on crossing the disc. The term in $C_2$ comes from the homogeneous part of the equation so we normally set it equal to zero. It would be necessary  in order to confront a specified boundary condition. The only quadrupole type solution is with $C_1=0$, which gives only a purely azimuthal field, possibly discontinuous at the disc
 
 When the outflow and lag functions have the forms of Eqns.~\ref{eq:FFvelocity}, rather than being constant, the solution for the field  becomes 
  \bea
b_R&=&\frac{e^{(2\delta-\alpha)T}}{rw(\zeta)}K\tilde A_\Phi(\zeta),\nonumber\\
b_\Phi&=& -\frac{e^{(2\delta-\alpha)T}}{rw(\zeta)}\big (\zeta v_1(1+\frac{K\zeta}{w(\zeta)})-\frac{Kv_1}{w_1}(\zeta+\frac{w_o}{w_1}\ln{w(\zeta)})+KC_2/C_1\big )\tilde A_\Phi(\zeta),\label{eq:FFlag+out}\\
b_Z&=& \frac{e^{(2\delta-\alpha)T}}{r}(1+\frac{K}{w}\zeta)\tilde A_\Phi(\zeta).\nonumber
\eea
The azimuthal vector potential has the  {\it power law} form
\be
\tilde A_\Phi=C_1e^{-\frac{K}{w_1}\ln{w(\zeta)} }.\label{eq:FFAphi}
\ee
 In these formulae $w_1\ne 0$ and,  if $w_o=0$, there is a logarithmic singularity at $\zeta=0$. 
 Once again the general behaviour is   that of a rising, widening, spiral as was first seen in \cite{HI2016}.  On crossing the equator, we might expect $w_o$ and $v_1$ to change sign, but not $w_1$ so that $w$ changes sign. Hence $b_R$, $b_\phi$ change sign but $b_z$ does not, implying dipole behaviour as in the constant velocity case.  We intend the real part of the logarithm in these expressions.

\begin{figure}{}
\rotatebox{0}{\scalebox{0.9} 
{\includegraphics{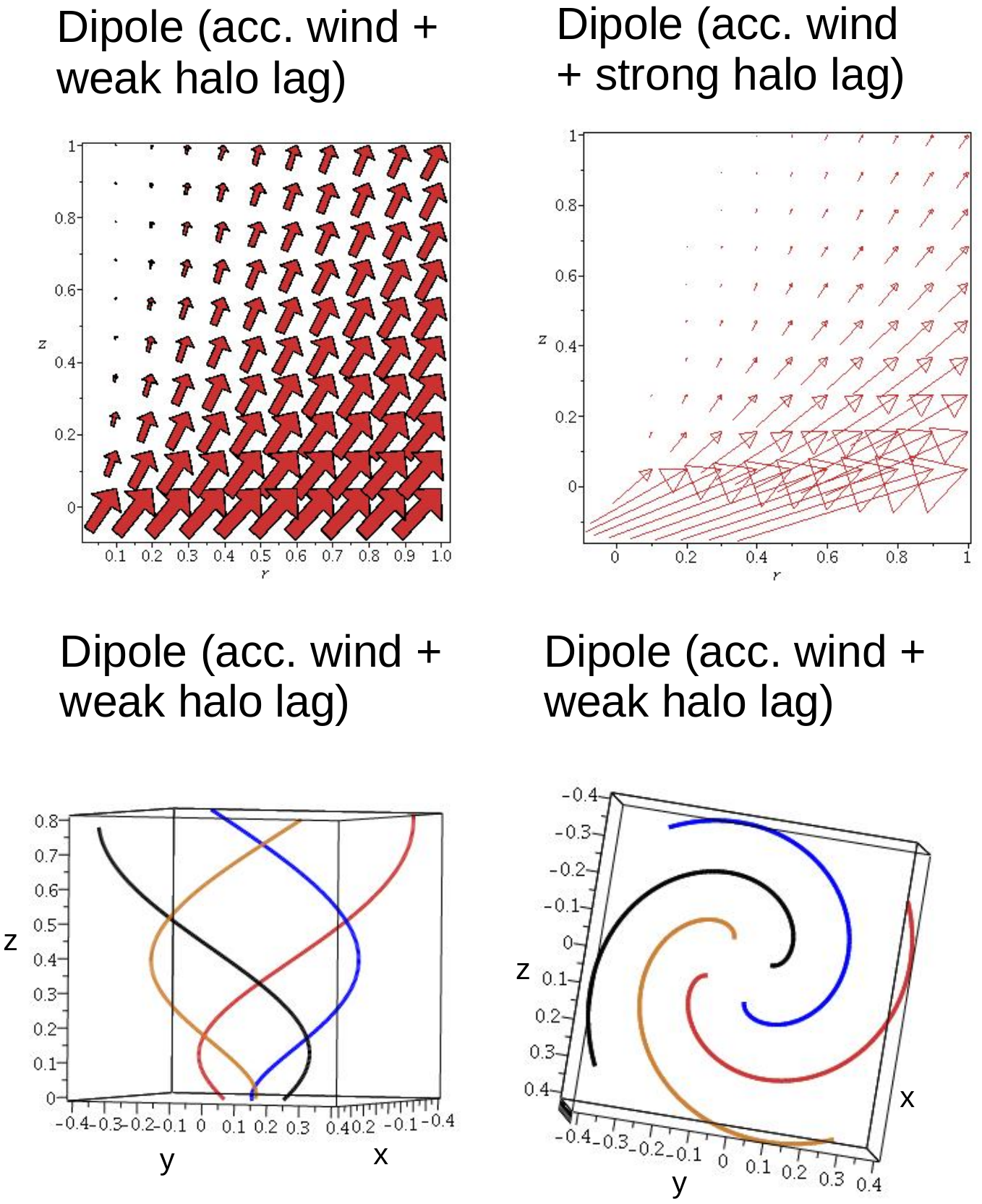}}}
\caption{In the bottom row we show a cluster of 3D field lines for the parameters $\{v_1,w_o,w_1,K,C_1,C_2\}=\{0.15,1.0,0.1,1,1,0\}$. The field lines all pass through $r_o=0.25$, $z_o=0.25$ and through  the cardinal points on the circle of radius $r_o$. The side view is on the left and the top view is on the right. There is no true quadrupole as can be seen from Eqns.~\ref{eq:FFlag+out} and \ref{eq:FFAphi}. The plot in the $\{r,z\}$ plane at upper left is for the same parameters. The plot at upper right shows the effect of a strong halo rotational lag with parameters $\{1,0.25,0.25,1, 1,0\}$. The arrows correspond to values at their head.}
\label{fig:analyticFF}
\end{figure}

 We append a relatively generic example  in Figure~\ref{fig:analyticFF} in order to fix ideas. Depending mainly on the relative size of the halo lag  $v_1$ relative to the wind acceleration,  $w_1$ the spirals are flatter when the ratio is large and more extended when the ratio is small. Recall that the Unit  in the figures is the disc radius. The constant wind velocity merely convects boundary conditions as we have seen before. This is because it does not contribute to a differential twisting of the field.
 
 
 
 The  `X type' field behaviour is  found in projection as a characteristic of rising, spiralling fields. It will be at an angle close to the disc when the halo lag rate is dominant and closer to the galactic axis when the wind acceleration is stronger. A strong wind acceleration case (not shown) has almost straight  field lines in the halo rather than flatly spiralling field lines. This is very similar to the quadrupole $\alpha_h$ behaviour. 
 
 
We summarize this section by comparing it  to the pure $\alpha_h$ dynamo of Sect.~\ref{sect:aux11}.  We have found widening, spiralling, halo structure in both limits as elaborated below. 

The pure $\alpha_h$  sub scale turbulent average dynamo produces tightly wound spirals near the axis of the galaxy, where they are unlikely to be observed.  The large radius spirals are very slow and  the projections into the $\{r,z\}$ plane show `X type' topology  only near the disc for the quadrupole solution.   The dipole  $\alpha_h$ solution shows a rapid decline in field strength with height in the azimuthal and vertical field components when the lag and wind are weak. 

The macroscopic flux freezing scale invariant dynamo of the present section produces  rising spiral magnetic fields  with large polar angle, when the halo rotational  lag velocity dominates the wind velocity.   The X type topology is then projected into the $\{r,z\}$ plane in a well defined low latitude sector. The wind dominated case shown at upper left gives a more typical X field topology. There is also scope for varying  the sign of the  Faraday depth above the disc  when the spirals are rapidly turning inside the galactic halo. This is especially likely when the halo rotational lag dominates.  



 Referring once more to \cite{KIS2020}, The flux freezing `dynamo' alone can not give both a high latitude X field  and a parallel field in the disc. However it shows clearly the association between the X field  topology and rising magnetic spirals. The wind dominated field is a natural explanation for the observed magnetic topology in NGC~3044 (see Figure~\ref{fig:alphaHalpha}).
 
 The C band image of NGC~4631 indicates similar poloidal behaviour, but the multiple sign changes in RM in this galaxy suggests  the presence of non axially symmetric spiral arms \citep[see e.g.][]{WHIM-P2019} in the halo.

\newpage
 \subsection{Latitude dependence of sub scale turbulence and rotational halo lag}
 
\label{sect:aux3}

We turn  to consider a sub scale turbulent dynamo varying with latitude and acted on by a rotationally  lagging halo. That is, we allow the auxiliary  turbulent quantity $\tilde\alpha_h$ to be a function of $\zeta\equiv z/r$.
By concentrating the {\it scaled} sub scale turbulent helicity to the galactic axis  in this way, we  imitate the effect that an AGN may have on the global galactic magnetic field. The effects of an axial wind must be left to another work.

The suggestion of an  axially symmetric `AGN dynamo' assumes that the AGN axis is approximately aligned with the axis of the galaxy. In galaxies where this is not the case we can expect more  asymmetric galactic magnetic fields, at least near the nucleus.  It is not necessary for a galaxy to have  an axial concentration of  magnetic field, such as may be produced by an AGN, in order to display a large scale coherent magnetic field. We have seen this in previous sections, especially with the quadrupole mode where the field may be tightly wound about the axis.

If we choose $\tilde\alpha_h$ to be linear in the tangent of the latitude, $\zeta$,  according to 
\be
\tilde\alpha_h=\alpha_{h1}\zeta,\label{eq:auxvariable}
\ee
then we must use $T$ as defined in Eqn.~\ref{eq:invariants}. Consequently,  the constant $\tilde\alpha_{h1}$ no longer vanishes from the equations for the vector potential.

We have, with Eqn.~\ref{eq:auxvariable}, assumed (recalling Eqn.~\ref{eq:helicity}) that the  physical sub scale turbulent helicity  increases linearly with height in the halo.  However the scale invariant solution may be expected to react to the latitude dependence in $\tilde\alpha_h$. The linear form satisfies the antisymmetry across the disc \citep[e.g.][]{KF2015}.  

This  form for $\tilde\alpha_h$ parallels the assumption regarding the form of the halo rotational lag (see Eqn.~\ref{eq:v}) relative to a pattern frame. In fact if we assume that our calculation is done in a uniformly rotating ($\Omega$) pattern frame, we can associate $2\Omega r$ with the base hydrodynamic vorticity  to which we add the value (Eqn.~\ref{eq:auxvariable}).

Eqns.~\ref{eq:dynamo1} for the components of the vector potential become,  when $\tilde\alpha_h$ varies generally with $\zeta$, 
\bea
K\tilde A_R&=&-\tilde\alpha_h(\zeta)\tilde A_\Phi'+v\tilde A_\Phi-\zeta v\tilde A_\Phi',\nonumber\\
K\tilde A_\Phi&=& \tilde\alpha_h(\zeta)(\tilde A_R'+\zeta\tilde A_Z'),\label{eq:varalpha1}\\
K\tilde A_Z&=& \tilde\alpha_h(\zeta)(\tilde A_\Phi-\zeta\tilde A_\Phi')+v\tilde A_\Phi' ,\nonumber
\eea
and these combine to give the equation for the azimuthal vector component. Once this equation is solved the other components and the magnetic field components follow. The equation  to be solved takes the general form
\be
(1+\zeta^2)\tilde A_\Phi''+\frac{d\ln{\tilde\alpha_h(\zeta)}}{d\zeta}(1+\zeta^2)\tilde A_\Phi'+(\frac{K^2}{\tilde\alpha_h(\zeta)^2}-\zeta\frac{d\ln{\tilde\alpha_h(\zeta)}}{d\zeta}-v')\tilde A_\Phi=0.\label{eq:varalpha2}
\ee



Using Eqns.~\ref{eq:auxvariable} and Eqn.~\ref{eq:v}, this  last  equation simplifies to 
\be
(1+\zeta^2)\tilde A_\Phi''+(\zeta+\frac{1}{\zeta})\tilde A_\Phi'+(\frac{K^2}{\alpha_{h1}^2\zeta^2} -1 +v_1)\tilde A_\Phi=0.\label{eq:working2}
\ee 

This equation  has a logarithmic singular point at $\zeta=0$. However in the limit $\zeta\rightarrow 0$ the solution is proportional to $\zeta^p$ where $p^2=-K^2/\alpha_{h1}^2$. Consequently the singularity is an oscillation in $\ln{\zeta}$  and is easily recognized as   $\alpha_{h1}\rightarrow 0$. We observe in our examples below that this oscillation is associated with the sub scale turbulence becoming dominant, as it extends to smaller scales.
 
 The  magnetic field  is no longer invariant under a change in sign of $\zeta$  according to Eqn.~\ref{eq:working2} because of the term in $v_1$,  which changes sign on crossing the disc. The sign of $\alpha_{h1}$ is irrelevant as one might expect because only the magnitude of the vorticity should produce the alpha dynamo. 
 
 
  A solution of Eqn.~\ref{eq:working2} at positive $\zeta$ can be pasted into the negative region (keeping the sign of $\tilde A_\Phi$ unchanged, but remembering to change the sign of $v_1$) to establish symmetry.  This yields, by Eqns.~\ref{eq:varalpha1} and the equations for ${\bf b}$,  a change in sign of $b_R$ and $b_\Phi$ but not in $b_Z$. 
 This would  construct  local dipole boundary conditions, but a quadrupole boundary follows by changing the sign of $\tilde A_\Phi$.
  However  the  oscillations in sign near the disc  due to small scale turbulence confuse the issue. These fluctuations become smoothed to a coherent quadrupole for  the $\{0,1\}$ solution, and to a coherent dipole for the $\{1,0\}$ solution when $v_1=\alpha_{h1}=O(1)$.

We observe, from Eqns.~\ref{eq:velocity}  and \ref{eq:helicity}, that if  indeed $\tilde\alpha_h=\alpha_{h1}\zeta$, {\it and} $v=-v_1\zeta$, then 
\be
v_s=-\delta rv_1\zeta e^{(-\alpha)T}~~~~~~\alpha_h=\alpha_{h1}\zeta e^{(-\alpha T)}.\label{eq:vaux2}
\ee
 It is convenient to choose $\delta r$ to be the actual lag rate so that $v_1=O(1)$.  This means that the value of $\alpha_{h1}$ is likely to be also $O(1)$.

The magnetic field components take the same form  as in Sect.~\ref{sect:aux1} except  for $\bar b_\Phi$. Eqn.~\ref{eq:bPhi} becomes, using the expression for $\bar A_\Phi$  in Eqn.~\ref{eq:varalpha1} divided by $\tilde\alpha_h(\zeta)$,
\be
\tilde b_\Phi=\frac{K\tilde A_\Phi}{\tilde\alpha_h(\zeta)}.\label{eq:bPhi2}
\ee
Recalling the linear form $\tilde\alpha_h$ (Eqn.~\ref{eq:auxvariable}) we see that the sign of $\alpha_{h1}$ affects the sign of $\tilde b_\Phi$ as one expects from helical (vortical) turbulence, but it does not affect the magnetic topology.

\subsubsection{ Examples}
\label{sect:variablealphasolutions}

The two independent solutions of Eqn.~\ref{eq:working2}  ($\{C_1,C_2\}=\{0,1\}$ or $\{1,0\}$) are complex conjugates. We choose the real part of the $\{0,1\}$ solution and the imaginary part of the $\{1,0\}$ solution.  This is because these choices best fit the expected quadrupole and dipole boundary conditions.

The oscillatory logarithmic singularity at small $\zeta$  is quite obvious  in the solutions for small $\alpha_{h1}$ near the disc for small values of $\alpha_{h1}$. This is   formally  because whenever $K^2/\alpha_{h1}^2$ dominates the factor of $\tilde A_\phi$ in Eqn.~\ref{eq:working2}, the oscillatory approximation extends to larger $\zeta$.  However $\zeta$ must still be less than one (polar angle greater than $45^\circ$), so that the oscillations occur near the galactic disc. 
  Physically, this is of interest because the $\alpha_h$ term in Eqn.~\ref{eq:dynamo1} is due to sub scale turbulence. As this term becomes dominant at smaller scales (i.e. smaller $z$ at fixed $r$) , we can expect turbulence to manifest itself in the magnetic field.

\begin{figure}{}
\rotatebox{0}{\scalebox{1.0} 
{\includegraphics{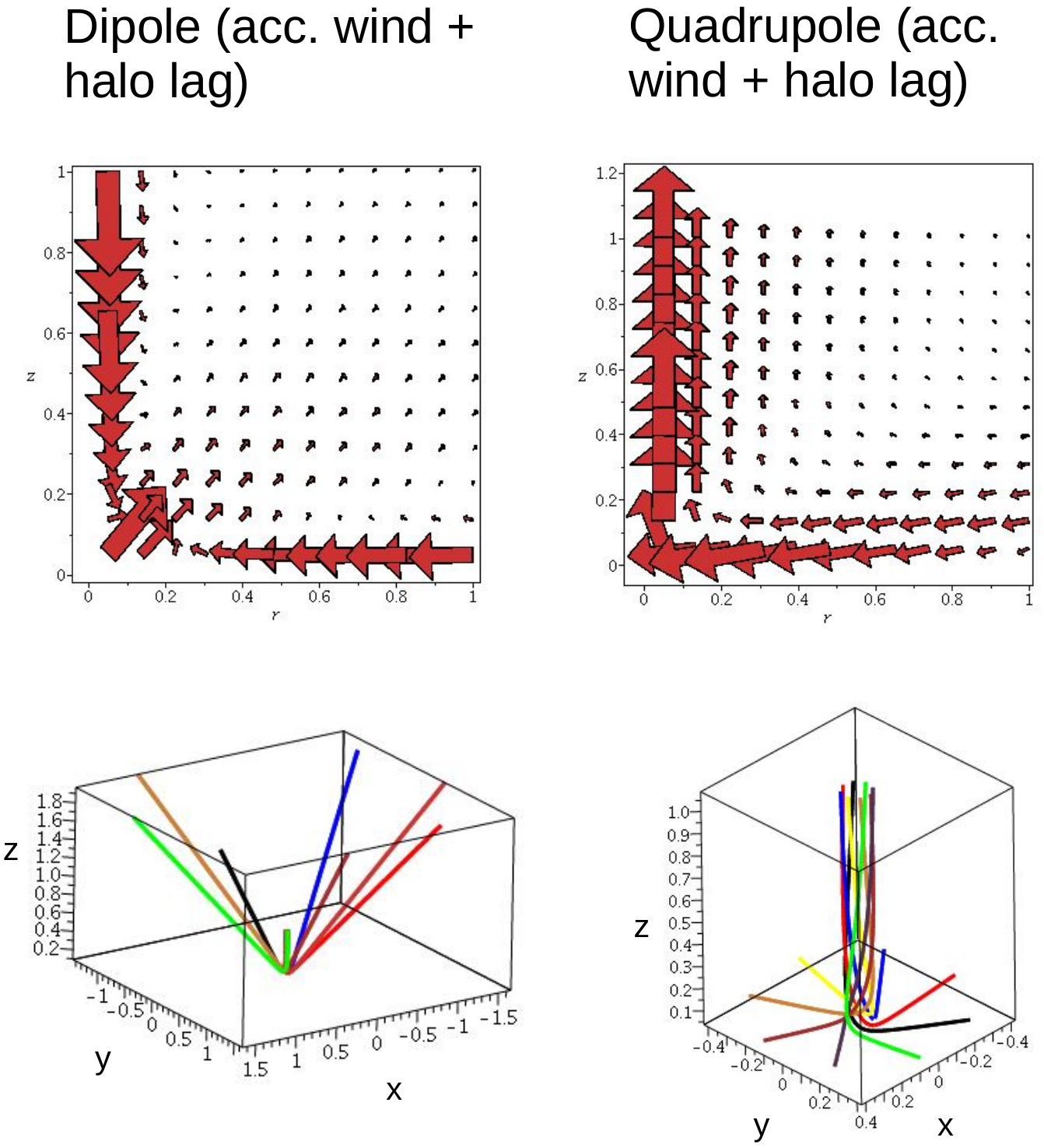}}}
\caption{In the left column, we show the dipole with parameters  $\{v_1,\alpha_{h1},K,r,z,C_1,C_2\}=\{1,1,1,r,z,1,0\}$.  
  The $\{r,z\}$ plane is on the top and a cluster of field lines is on the bottom. All field lines in this figure pass through $r=0.05$ and $z=0.1$ and are distributed in steps of $\pi/4$ in azimuth. The  $\{r,z\}$ cuts must be imagined rotated about the axis and the field line plots are filled in by similar field lines. The right column illustrates, in the same manner, the quadrupole with parameters $\{1,1,1,r,z,0,1\}$. }
\label{fig:varalphalag1}
\end{figure}

The $\{r,z\}$ plane for the $\{0,1\}$ (quadrupole) solution is shown at upper right in Figure~\ref{fig:varalphalag1} and some corresponding field lines are shown at lower right.   The field lines are constrained to the same scales on each axis.
The images in the left column of the figure are for the $\{1,0\}$ (dipole) solution. Strong field lines descend  rapidly near the axis  and then rise.  The `X type' projected field lines are confined to a  range of latitude near $45^\circ$ Latitude.

\begin{figure}{}
\rotatebox{0}{\scalebox{1.0} 
{\includegraphics{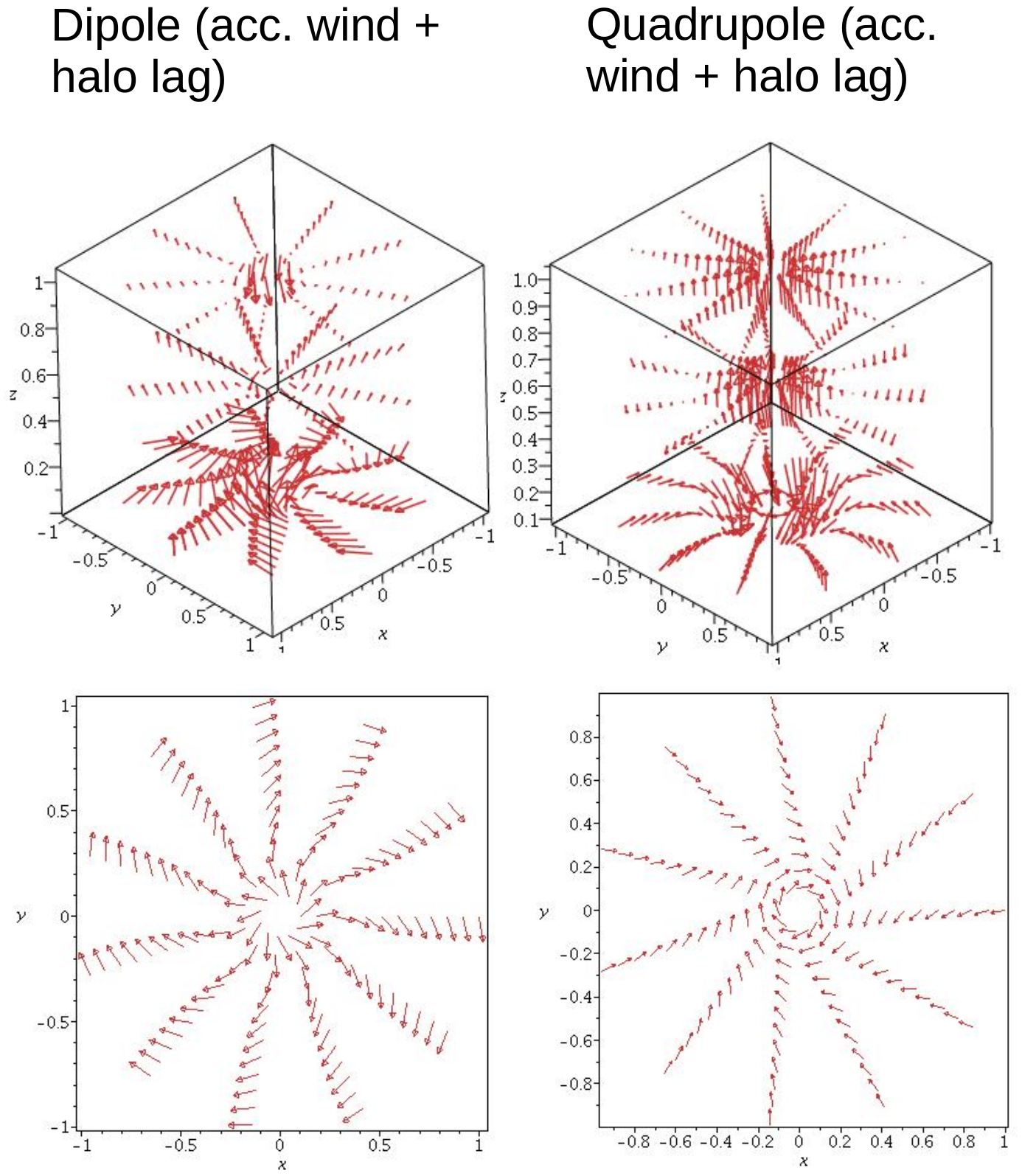}}}
\caption{The figure at upper left is the 3d version of the case shown in the left column of figure (\ref{fig:varalphalag1}). The parameters are the same $\{v_1,\alpha_{h1},K,r,phi,z,C_1,C_2\}=\{1,1,1,r,phi,z,1,0\}$. The right column has the parameter set $\{1,1,1,r,phi,z,0,1\}$ and corresponds to the right column in figure (\ref{fig:varalphalag1}).
The lower panels show an $\{r,\phi\}$  toroidal projection at height $z=0.25$ in each case for the same prameters.}
\label{fig:varalphalag2}
\end{figure}

One should remember that in all mages of the $\{r,z\}$ plane the diagonal, apparently rectilinear magnetic vectors,  are actually turning in the $\{r,\phi\}$ plane. Thus they represent rising  or falling spirals as confirmed in Figure~\ref{fig:varalphalag2}. The twist rate differs depending on the parameters and frequently varies with height above the disc and radius in the disc. 

The quadrupole behaviour  is strongest parallel to the axis and also parallel to the equator. The transition in field direction occurs near $45^\circ$ and removes any X type topology in favour of a nearly vertical field at large latitude.  The corresponding field lines in the lower right panel illustrate that the field near the axis is nearly vertical with small toroidal field. The upper right panel shows how field lines starting at higher latitude fill the space of the lower panel.

The dipole behaviour in the left column of Figure~\ref{fig:varalphalag1} has a descending strong vertical field near the axis that rises almost rectilinearly, as seen in the field lines in the lower panel. These produce the X field topology near $45^\circ$. The toroidal plot in Figure~\ref{fig:varalphalag2} indicates only a very small azimuthal field compared to the quadrupole. These figures are for nearly equal lag velocity and sub scale turbulent velocity.

Figure~\ref{fig:varalphalag2} shows, at upper right, the $3d$ version of the $\{r,z\}$ cut at upper right in Figure~\ref{fig:varalphalag1}. The  toroidal cut of the same field is shown at lower right.
The field lines  descend from the halo at large radius and then rise again at latitudes greater than $\pi/4$.  The very slight toroidal field near the axis compared to the vertical field is shown in the lower right panel.


At upper left in Figure~\ref{fig:varalphalag2}, we have the $3d$ image of the dipole with the same parameters.  The field lines descend while turning  slowly and subsequently rise rectilinearly at larger radius. This develops {in a self similar fashion from all heights.}
 The  transition region contains a projected `X field' to be present mainly near $45^\circ$. The azimuthal components of the vectors change sign with radius.



\begin{figure}{}
\rotatebox{0}{\scalebox{1.0} 
{\includegraphics{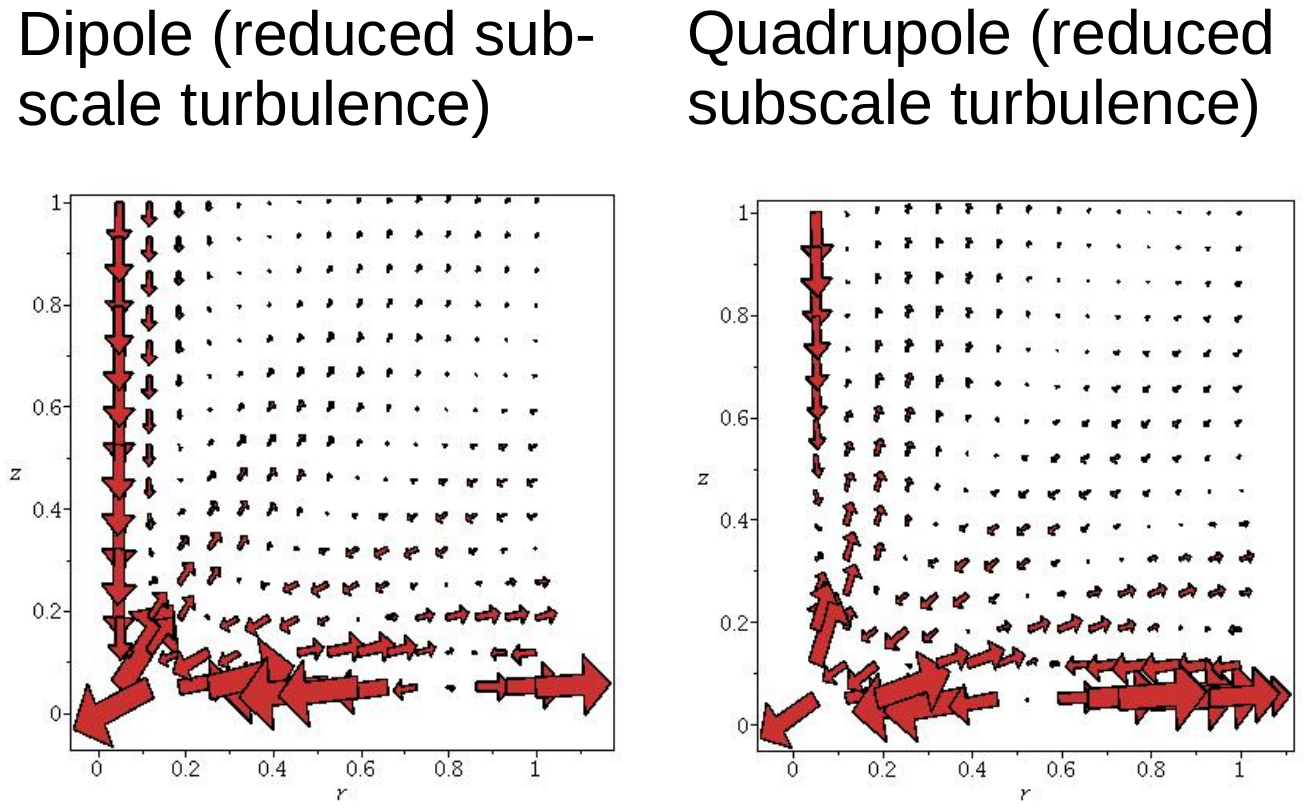}}}
\caption{ The image at left shows the $\{r,z\}$ field line vectors for the dipole case 
$\{v_1,\alpha_{h1},K, r,\phi,z,C_1,C_2\}=\{1,0.25,1,,r,\phi,z,1,0\}$. At right we have the quadrupole field line vectors with parameter set $\{1,0.25,1,r,\phi,z,0,1\}$.
 These reveal the onset of galactic disc small scale magnetic turbulence. }
\label{fig:varalphalag3}
\end{figure}

In  Figure~\ref{fig:varalphalag3}, we  show the effects of reducing the scale and strength of the sub scale turbulence. The halo rotational lag has been kept constant but $\tilde\alpha_{h1}=0.25$ in both the quadrupole and  dipole panels.  Both modes now show X field topology in two narrow but separated sectors. The quadrupole actually shows three X field sectors, but one is very close to the equator of the galaxy. The more interesting development is the increasing turbulence in the disc of the galaxy for both modes. This tendency increases in latitude extent as $\tilde\alpha_{h1}$ decreases further.

This small scale turbulence is to be expected when the alpha dynamo dominates.  Its appearance actually represents a confirmation of the form for this sub scale dynamo, as used in the classical Eqn.~\ref{eq:dynamo1}.



An interesting result of this section is the appearance in the dipole of strong axial magnetic field, together with X field topology that is intermediate between the axis and the equator.  Although we have not included a `jet' or a nuclear `wind' directly, the dipole produces an axial field that we  know from previous sections would not be changed by a constant vertical outflow velocity required in a jet or wind. This is achieved this by assuming a strong source of scaled sub scale vorticity on the galactic axis. This can be produced by an AGN, that is, by a super massive black hole or a burst of star formation. {\it Such nuclear flows may therefore be an important contributor to the global galactic magnetic field.} 

If jets  or nuclear star bursts do contribute to the coherent galactic magnetic field, we can expect a correlation between the presence of a galactic AGN and  the  strength of the magnetic  field. In fact the  disc field is also strong and turbulent (see Figure~\ref{fig:varalphalag3}). Both the correlation and the turbulence would indicate an important sub scale dynamo. Similar global magnetic topology can be produced without an AGN  but the strong axial field  will be absent. This prediction may be readily tested for galaxies where the AGN axis and galactic axis coincide.

\begin{figure}{}
  \begin{tabular}{cc} 
    \hspace{-0.4truein}
\rotatebox{0}{\scalebox{0.6} 
  {\includegraphics{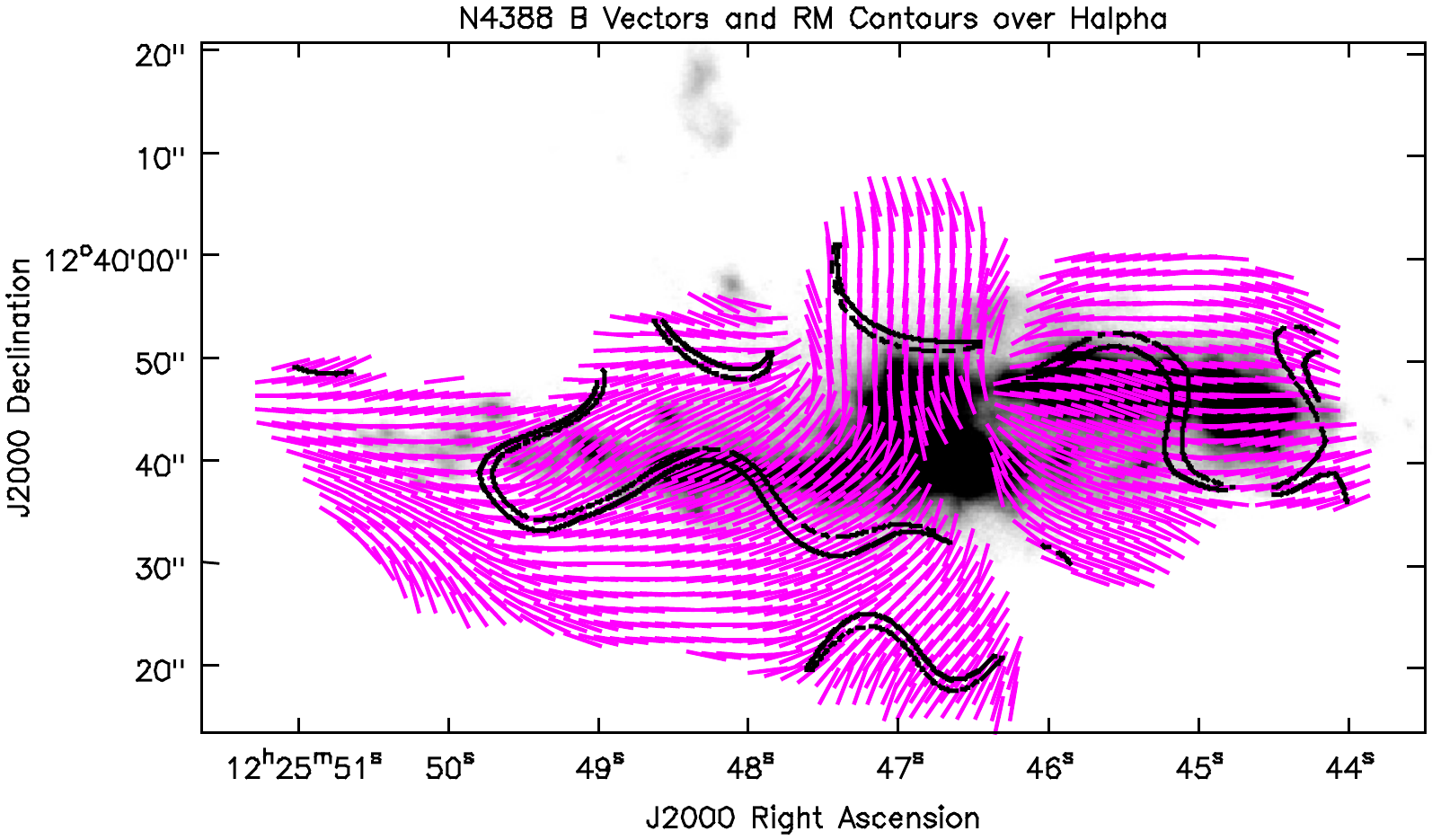}}}&\hspace{-0.2truein}
\rotatebox{0}{\scalebox{0.4} 
  {\includegraphics{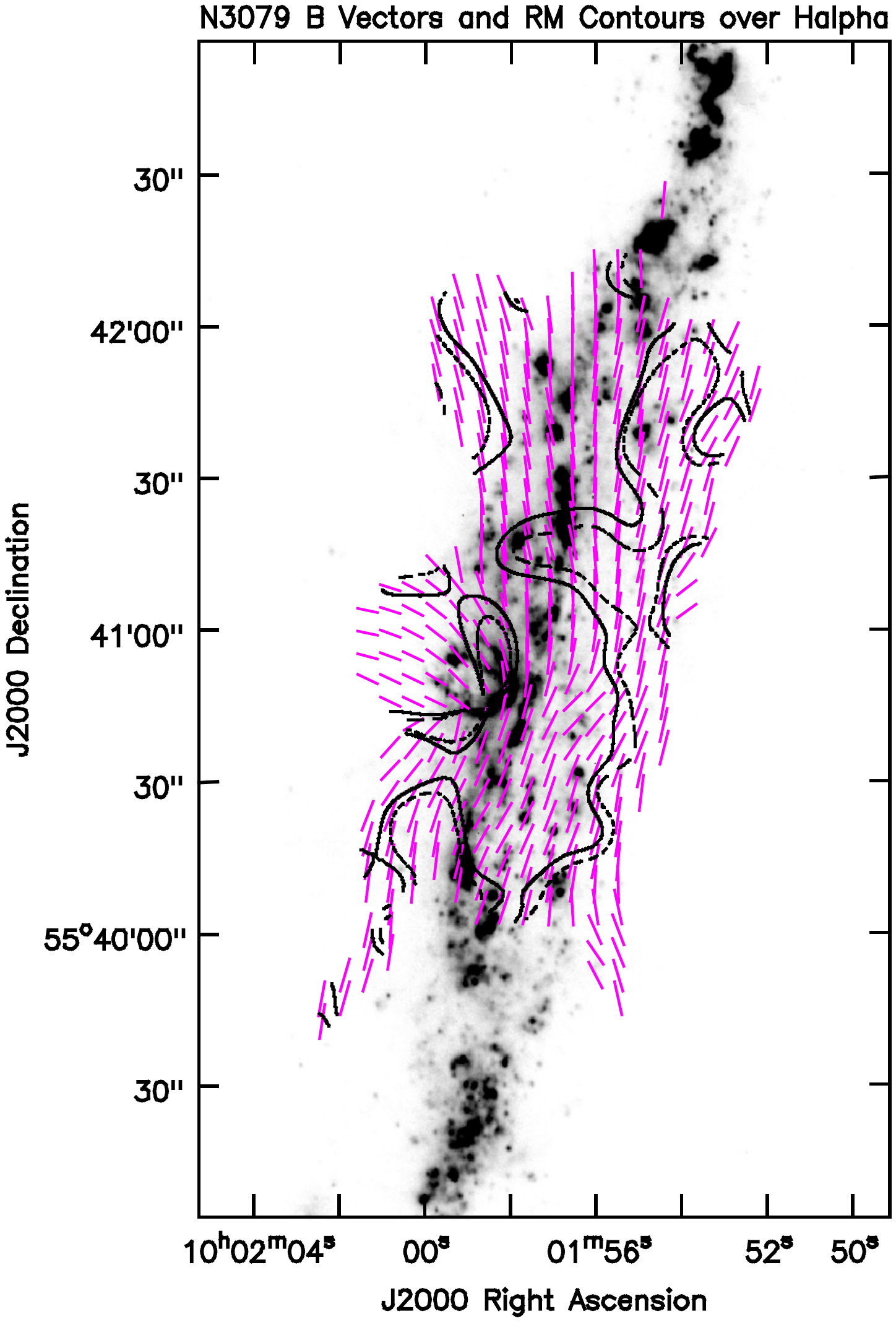}\hfill}}
\end{tabular}
\caption{NGC~4388 (left) and NGC~3079 (right) with H$\alpha$ emission shown in greyscale. 
B vectors, corrected for Faraday rotation, are shown in magenta. RM contours have been set only for the purpose of indicating where there is a sign change from positive to negative or vice versa and are present to help guide the eye.
The contours are set to +10 (solid curves) and -10 (dashed curves) rad m$^{-2}$ for NGC~4388 and +15 (solid curves) and -15 (dashed curves) rad m$^{-2}$ for NGC~3079. 
See \protect\cite{KIS2020} for more explanation of the polarization data and \protect\cite{varg18} for information on the H$\alpha$ image. 
All data have been downloaded from queensu.ca/changes. }
\label{fig:jetscoherent}
\end{figure}

Perhaps the best candidates from CHANG-ES results are NGC~4388 and NGC~3079.  Recent images of these two galaxies are shown in Figure~\ref{fig:jetscoherent}.   There is indeed a correlation between the strength of the coherent galactic polarization and the presence of AGN activity as a nuclear wind.
The RM sign changes in NGC~4388 tend to indicate spiralling fields above the disc.  This is less clear in NGC~3079 but also present.  NGC~4388 shows rapid changes in field direction in the disc region, which may be turbulent.
 


\newpage

\section{Discussion }

Our intention in this paper has been to explore scale invariant, axially symmetric, time dependent magnetic dynamos. We have proceeded by isolating and combining the turbulent, $\alpha/\Omega$ and flux freezing `dynamos'. The diffusion has been neglected  under defined conditions. Formally this  assumes infinite conductivity. Excess flux can be removed from the galaxy by a disc wind with a gradient.



\subsection{Sub scale turbulent dynamo only}

In Sect.~\ref{sect:aux11}, we studied the $\alpha_h$ dynamo alone and displayed the results for the quadrupole and dipole in Figure~\ref{fig:alphaeffect}. There are very few parameters in this part because we chose $\tilde\alpha_h$ to be constant with latitude. This allowed us to remove it from the equations by placing it in the time dependence and as a factor in the velocity. 

 In Figure~\ref{fig:alphaeffect}, one sees  in the dipole a very slow spiral twist except near the axis, as  is shown in the lower row by an axial field line. There is also a characteristic oscillation in the direction of the azimuthal field  with height  in the dipole. The quadrupole  on the right has slowly spiralling field lines as indicated in the lower row of the figure.  Clusters of these field lines would project only weakly as `X type' topology above the plane as indicated in Figures~\ref{fig:fieldlinesalpha} and \ref{fig:alphaHalpha}. Notice that  no strong vertical field is produced near the axis of the galaxy and the turbulent dynamo produces only weak,nearly vertical, X field.

 It should be remarked that our model  generally has no absolute measure of the strength of the magnetic field unless we have a measure of such a field at a given time. We can say that we expect $\tilde\alpha_h\delta r\le\approx 0.1$ if $\delta r$ is comparable to the rotation speed of the disc. The growth time scale is then $\approx 1/(\delta \tilde\alpha_{h1})$, that is, perhaps equal to $10$ rotation periods.  The period would be measured at the start of the flat rotation curve of the galaxy.  
 

\subsection{Turbulent dynamo plus rotational  lag}

In Sect.~\ref{sect:helicitypluslag}, we added a halo lag to the $\alpha_h$ dynamo. The lag is a form of the macroscopic $\alpha/\Omega$ dynamo. The rotational lag is linear with height and we take the lag rate in our examples to be typically $dv_\phi/dz = -\alpha_d\delta\approx 10-20~ km/sec/kpc$. In Figure~\ref{fig:alphalag1}, we show clusters of field lines and $3d$ vector plots for the quadrupole and dipole. Figure(\ref{fig:alphalag2}) shows ($\{r,z\}$ and $\{r,\phi\}$) cuts. The rotational lag increases the twist of the quadrupole relative to the zero lag case, but the poloidal field is not greatly changed. The dipole has been more strongly affected. It shows a stronger X field topology and continues to show what would be a sign change in the RM with height.





\subsection{Turbulent dynamo plus wind and lag}

In Sect.~\ref{sect:alphawind}, we have added an outflow/inflow to the turbulent dynamo with rotational lag.  Figure (\ref{fig:alphawind1}) shows, for both the quadrupole and the dipole, the strong and weak wind, zero  rotational halo lag, $\{r,z\}$ vectors.  It is clear that the strong wind gradient `straightens' (more vertical ) the spirals in both cases and migrates the field to greater heights. The dipole  field lines are rendered almost vertical. 
Moderate wind disc velocity and gradient enhances the X field topology.


Figure~\ref{fig:alphawind2} shows sample field lines for the quadrupole on the left and the dipole on the right.   The upper row has no rotational halo lag
but moderate wind parameters corresponding to the upper row in Figure~\ref{fig:alphawind1}. The lower row adds a typical rotational halo lag and we 
see the field lines spiralling at lower heights. 


In Figure~\ref{fig:alphawind3}, we have shown the peculiar polarization of the galaxy NGC~4013 as a possible example  of quadrupole turbulent field distorted by a lagging halo.


\subsection {Flux Freezing}

In Sect.~\ref{sect:aux2}, we have focused on the effect of pure flux freezing scale invariance. There is no current turbulent dynamo so that at some earlier time there must be  a seed field. The field will eventually decay according to the discussion in Sect.~\ref{sect:time}, depending on global constants. The reason for dwelling on this  rather restricted `dynamo' is that : (a)~~it is wholly analytic and ~~(b) it creates at least a temporary magnetic field similar to that of a real dynamo. An analytic solution  also exists when the magnetic field is not axially symmetric but rather forms magnetic spiral arms. We do not pursue this in this paper. 

The results of this section are summarized in Eqns.~\ref{eq:FFcstvel} and \ref{eq:FFlag+out}. However Figure~\ref{fig:analyticFF}  
shows a cluster of field lines  for the dipole with moderate acceleration and lag compared to the outflow velocity.  This shows beautifully the rising spiral topology of the magnetic field.  The $\{r,z\}$ cuts  in the upper row are for the same parameters on the left. The X field is strongly present. When there is  strong rotational lag and weak wind, the low latitude X field at upper right is found.


The forms of the explicit equations for the magnetic field
are noteworthy for their description of the $\{r,z\}$ dependence.  In the constant {\it scaled} velocity  (there is a linear radial dependence for the physical velocity) case we find an exponential decline as a function of $-z/(rw)$. The scale height is therefore $wr$ and increases both with $w$ and $r$. This produces a higher magnetic halo at larger radius and with stronger wind. This can be compared to the CHANG-ES paper conclusions in this regard \citep{KIW2018}. 

The behaviour found here  affords an explanation of the correlation between scale height and diameter found in \cite{KIW2018},  provided that galaxies  scale similarly as well as self-similarly.  A superposition of the coherent magnetic fields in CHANG-ES galaxies \citep{KIS2020} encourages this view.

The solution with variable  wind and halo rotational lag is more complicated, but varies in the halo essentially with the power $(w_o+w_1z/r)^{-K/w_1}$. For accelerating outflow, when $w_o$ and $w_1$ are positive, the magnetic halo has a similar structure to that given by  constant velocity although it is not exponential. The halo structure in $z$ evidently can be more subtle with $w_1<0$.

\subsection{High latitude concentration of turbulence plus rotational lag}

Sect.~\ref{sect:aux3} illustrates  how  the sub scale turbulent dynamo might appear when the turbulence is strong near the galactic axis. Figure~\ref{fig:varalphalag1} shows an $\{r,z\}$ cut at upper right for the quadrupole  and corresponding field lines at lower right. The concentration of field to the axis is evident, but there is no X field. The left column shows the dipole for the same parameters. The field is again very strong at height and there is an X field at $45^\circ$.

 Almost unique  to this case is the dominant axial poloidal field. Something similar is found with constant $\tilde\alpha_h$ but it is not as pronounced. The `X field' region is extensive in the dipole. This  morphology of X field and `jet' axial field is quite distinct from the other examples in this paper. 

The  upper $3d$  and $\{r,z\}$ images in Figure~\ref{fig:varalphalag2} have the same parameters as the upper panels in Figure~\ref{fig:varalphalag1}. These confirm the structure found in the $\{r,z\}$ plane.  The  lower images show the field lines for the same parameters.  The quadrupole shows the extreme axial field.

Figure~\ref{fig:varalphalag3} shows an unexpected result of reducing the amplitude and vertical scale of the sub scale turbulence while it becomes dominant.  The turbulence becomes visible in the disc of the galaxy.

This section led us to suggest that the AGN  (jet or nuclear wind) may have a part to play in the large scale coherent galactic magnetic field. We would predict isolated X field, strong parallel equatorial field, possibly turbulent, and strong field parallel to the axis as indicative of an `AGN  dynamo' with halo  rotational lag.


The only element missing in this study of the `parts' that make up an axially symmetric  scale invariant galactic dynamo is  diffusion. The {removal} of the increasing magnetic flux is accomplished here by either outflow or {implicitly} by a  globally conserved quantity. This was discussed in Sect.~\ref{sect:time}. We believe that this catalogue can help in identifying some of the processes at work in galactic magnetism. Magnetic arms have been discussed elsewhere and lead to quite  recognizable RM structures in the halo.

 \subsection{Relevant numerical work}

The recent numerical study, \cite{BZK2017}, is remarkable for its detailed simulation of what we refer to as the sub-scale turbulent dynamo. A widely distributed, naturally occurring, set of supernovae are used to pump energy and magnetic field at stellar scale into the interstellar medium. A numerically established, unmagnetized, spiral galaxy is the initial condition. In their figure {\it 2} the evolved magnetic field is shown as spiral in the disc but with no convincing `X field' in the halo. The field does not generally extend into the halo as far as indicated by observations. We conclude that additional lag or wind effects are required as we have indicated in this paper. There is a remarkable convergence in the conclusion stemming from two quite different methods.

In the paper \cite{PVB2020} the authors examine a suite of cosmological simulations to determine the outflow of magnetic flux from a disc galaxy into its halo and more distant surroundings. Their conclusion is similar to our own, in that coherent fields require coherent gas flows. However they also require sub-scale turbulence to amplify an initial seed field, to the point where coherent flow can operate. This turbulence operates even in the circum-galactic gas, just as we assume in Eqn.~\ref{eq:auxvariable}. 

Their Figure~8 shows the turbulence actually present in the gas and magnetic field. In our approach this turbulence must produce an electric field parallel to the magnetic field for amplification to occur \cite{SKR1966}. We have found such turbulence to appear at small $\zeta$, but with the turbulent velocity of Eqn.~ \ref{eq:auxvariable}, the energy spectrum is $\propto \zeta^3$, that is $k^{-3}$ with $k\propto \zeta^{-1}$. A Kolmogorov spectrum requires $\alpha_h\propto \zeta^{1/3}$ which changes the Eqn.~\ref{eq:varalpha2} slightly. This does not change the qualitative amplification ability of the turbulence.

{More recent papers, already cited  in the introduction, study the driving of the  galactic wind and the evolution of the magnetic field starting from early magneto genesis. The important conclusion from our perspective is the successful amplification of plausible primordial fields to those galactic fields observed today. This seems to require the action of a distributed turbulent ($\alpha/\Omega$) dynamo during galaxy formation. The wind  driving  has been shown to be associated with vigorous star formation and the resulting $\alpha_h$ turbulence, provided that the supernovae are inserted discretely. None of these results conflict fundamentally with the results of this and our earlier papers that take the scale invariant asymptotic view, i.e. the limiting consequence of all the physical interactions in the numerical developments is the scale invariant solution. }

There is a kind of consensus that the `X field' topology is due to the action of galactic outflow on a pre-existing field of stellar scale origin. It is easy to show that an extended wind in spherical geometry (e.g. \cite{RC1985}), centred on the nucleus of a galaxy  and acting on a pre-existing radial field component, will produce an `X field' globally given a realistic rotation law. This is most readily shown using an Lagrangian approach to the field evolution, but we must leave the demonstration to other work.

\section{Summary}

In this paper we have not attempted detailed modelling of any galaxy. We hope rather to motivate subsequent work in this direction by encouraging the simplifying assumption of scale invariance. To this end we have compared the contribution of various elements of the scale invariant dynamo to the polarization morphology of the CHANG-ES edge-on galaxies. Most  axially symmetric observational structures can be reproduced \citep[for nonaxial symmetry, see][]{WHIM-P2019} and the effects of different elements are sufficiently different that observations may eventually distinguish them.  Scale invariance also finds support in the stacked image of 28 rather different galaxies shown in figure 1 in \cite{KIS2020}. 
When each individual galaxy is spatially scaled to the largest galaxy size, the stacked image still reveals coherent magnetic field structures; this would not be present without a certain scale invariance.


\section{Acknowledgements}
{We thank an anonymous referee for hard work and helpful comments, and who found a gentle way of telling us what we should know.}


\label{lastpage}
\end{document}